%% file: wave.tex
\documentclass[a4paper,openany]{report}

\usepackage[T2A]{fontenc}
\usepackage[utf8]{inputenc}
\usepackage[english]{babel}
\usepackage{graphicx}

\author{\bfseries  ANDREY VASILYEV}
\title{\bfseries WAVE MECHANICS: BEHAVIOR OF A DISTRIBUTED 
ELECTRON CHARGE IN AN ATOM}
 
\date{Retired from State Optical Institute, Saint Petersburg, Russia\\
e-mail: andrey@wavemech.org}

\begin{document}

\maketitle

\begin{abstract}
\label{abstract}
A hypothesis is forwarded of the electron charge in an atom 
existing in a distributed form. To check it by methods of 
electrodynamics and mechanics (without invoking the formalism 
of quantum mechanics and the concepts of the wave function and 
of the operators), the potential, kinetic, and total energies 
were calculated for three states of the hydrogen atom, which 
were found to agree closely with the available experimental 
data. The second Chapter of the Paper offers additional 
assumptions concerning various scenarios of motion of elements
of the distributed electron charge which obey fully the laws 
of theoretical mechanics. The angular momentum of the 
ground-state hydrogen atom calculated in the frame of theoretical
mechanics is shown to coincide with the spin which is $\hbar/2$. 

\bigskip

{\bfseries Key words :}  quantum mechanics, electrodynamics, charge, 
potential, electric field, energy, angular momentum.
\end{abstract}

\tableofcontents

\input{ch1}

\input{ch2}

\end{document}

%% file: ch1.tex
\chapter{Spherical atom}
\label {ch1}

\section{Introduction}

Quantum mechanics permits one to calculate with a high 
accuracy all parameters of an electron in an atom, such 
as its energy, squared angular momentum, projection of 
the angular momentum on the $Z$ axis and others. The square 
of the wave function determines the probability for an 
electron to be at a certain point in space. It is possible, 
however, to study the behavior an electron in an atom in a more detailed 
way. One may invoke for this purpose methods developed in 
electrodynamics. Significantly, one can calculate many 
fundamental parameters of the electron in an atom without 
resorting to such strictly quantum-mechanical concepts as 
“wave function” or “operator”. It turns out also that such 
an approach provides in some cases a \textbf{more detailed} description 
of the behavior of an electron than it would be possible within 
the frame of quantum mechanics. This opens a way to understanding 
this behavior more closely. Accordingly, our basic goal here will 
be not determination of the parameters of an electron in an atom, 
which are actually well known, but rather an attempt at describing 
the electron behavior in greater detail. Now if the various 
parameters of the electron calculated in the context of 
electrodynamics are found to agree with experimental data and 
values derived from quantum mechanics, this may be considered as 
supporting the assumptions formulated below on electron behavior 
(or, in the case of disagreement, as refuting these ideas).

The first Chapter calculates the potential, kinetic, and total 
energies of the hydrogen atom for three states based on the above 
assumptions and invoking only the concepts of electrodynamics and 
theoretical mechanics.

The second Chapter calculates, likewise drawing only from the 
concepts of electrodynamics and theoretical mechanics, the angular 
momentum of the hydrogen atom in the ground state, which turned 
out to be $\hbar/2$, in excellent agreement with experimental data. 
We are going to conduct the relevant reasoning and calculations in 
the nonrelativistic approximation.

\section{Equation for the shape of a distributed electron charge 
in an atom}

Quantum mechanics assumes the electron to be a point charge of 
magnitude $(-e)$. A point charge at rest can be presented in the 
form of an expansion in a Fourier integral in scalar spatial 
harmonics of the {\itshape\bfseries {charge}}:
\begin{equation}
\label{1.1.1}
q(\mathbf{r}) = (-e)\delta(\mathbf{r}) = (-e)\int\limits_{-\infty}
^\infty f(\mathbf{k})e^{i\mathbf{kr}} d\mathbf{k},
\end{equation}
where $d\mathbf{k}=dk_xdk_ydk_z$ denotes integration over the 
three wave vectors $k_x$, $k_y$, $k_z$
of Cartesian coordinates. Here $k=2\pi/\lambda$, where  
$\lambda$ is the wavelength of the corresponding harmonic of 
{\itshape\bfseries {charge}}, 
is the wave number, and $\delta(\mathbf {r})$ is Dirac’s delta 
function. The spectral density $f(\mathbf{k})$ of expansion 
into the Fourier integral is defined in the following way: 
\begin{equation}
\label{1.1.2}
f(\mathbf{k})=\frac{1}{(2\pi)^3}\int\limits_{-\infty}^\infty \delta
(\mathbf{r})e^{-i\mathbf{kr}} d\mathbf{r},
\end{equation}
where  $d\mathbf{r}=dxdydz$ denotes integration over the three 
Cartesian axes.

Each harmonic $f(k_x)e^{ik_xx},$ $f(k_y)e^{ik_yy},$ $f(k_z)e^{ik_zz}$ 
in Eq. (\ref{1.1.1}) is essentially a wave at rest with wave numbers  
$k_x,$ $k_y,$ $k_z.$ These are, however, not electromagnetic 
but rather {\itshape\bfseries{charge waves.}}

If a point charge propagates with a constant velocity $\upsilon_x$ 
along the $x$ axis, it can be presented in the form of an expansion 
in moving harmonics:
\begin{equation}
\label{1.1.3}
q({\mathbf{r}},t)=(-e)\delta(\mathbf{r}-\mathbf{v}t)=(-e)
\int\limits_{-\infty}^\infty f(\mathbf{k})e^{i(\mathbf{kr}-wt)}d\mathbf{k}.
\end{equation}
Each harmonic $f(k_x)e^{i(k_xx-wt)}$ is a plane wave moving with a 
velocity $\upsilon_x$ along the $x$ axis. The frequency $w$ of 
all these waves is coupled to the wave number $k_x$ 
through the relation $k_x=w/\upsilon_x$. The $f(k_y)e^{ik_yy}$ and 
$f(k_z)e^{ik_zz}$ harmonics are, as before, waves at rest 
with wave numbers $k_y$ and $k_z$.

It should be stressed that the above decompositions {\itshape\bfseries
{are in no way }} expansions in de Broglie waves. These are nothing 
more than conventional expansions in spatial or spatial-temporal harmonics.

Significantly, expansion of an object in harmonics is not just a 
mathematical abstraction. Each harmonic is in actual fact a real 
physical component of a given process. For instance, when sound or
electromagnetic waves are decomposed into a spectrum each harmonic 
corresponding to a harmonic of the mathematical expansion is a real 
sound or electromagnetic wave exhibiting all the properties 
of a wave of this nature. The validity of this statement is not 
questioned any longer. Nevertheless, one could refer in this 
connection to the book by A. A. Kharkevich \cite{1} where this 
point is considered at length. In our case, we meet with the same 
situation in expanding a charge in separate harmonics of the kind 
of Eq. (\ref{1.1.3}), which represent in this particular case 
{\itshape\bfseries{charge waves.}}

Consider now a point electron that enters an atom and starts 
rotating about the nucleus. When an electron rotates about the 
nucleus, each {\itshape\bfseries{charge}} harmonic adding up 
to the charge of the electron must satisfy the periodicity 
conditions. Because the lengths of the {\itshape\bfseries{charge}} 
waves making up the electron charge are 
different, the shape of the point charge of the electron should 
somehow change. For this reason, the electron, while being 
point-like in free space, will have to “spread out” in rotation 
about the nucleus, i.e., assume another shape, different from 
point-like. Significantly, this will be a real “spread-out”, a 
real change in the shape of the point charge, rather than the 
probability of finding a point charge at a certain point in space. 
Said otherwise, we have here what can be called distribution of 
the electron charge in space.

Now what shape can assume a charge rotating about a nucleus? 
In other words, what will be the electron charge distribution 
in an atom? The Schr\"odinger equation permits one to calculate 
with a high precision the various parameters describing the 
state of an electron in an atom. This licenses us to assume 
that the charge of an electron rotating about a nucleus should 
take on the shape coinciding with that of an eigenfunction of 
Schr\"odinger’s equation, or of a function related in some way to 
eigenfunctions of the Schr\"odinger equation. The set of 
eigenfunctions of the Schr\"odinger equation is usually expressed 
in a spherical reference frame and is well known. Therefore, 
we are going to use in what follows the spherical coordinate 
system $r$, $\vartheta$, $\varphi$. The Cartesian system was 
used above only to make our reasoning more revealing.

What are the eigenfunctions of the Schr\"odinger equation 
among which one could look for functions describing the shape 
of a rotating charge? We note immediately that the shape of 
the distribution of this charge can be looked for only by 
invoking $S$ states of solutions of the Schr\"odinger equation. 
Indeed, charge conservation dictates that the total charge 
of a distribution rotating about a nucleus must be equal to 
the electron charge. Integration over the whole space of the 
$S$ states only yields a nonzero result. The integral over all 
other states gives zero. Therefore, only the $S$ state is capable 
of describing the presence of an electron in an atom. While the 
atom can contain, besides $S$, other charge states as well, they 
will be able only to modify the charge shape.

\medskip

Hence, {\bfseries\underline{in all situations}} the $S$ charge 
state  {\bfseries\underline{ is present}} in an atom.

The $S$ charge state {\bfseries\underline{ must be present}} in an 
atom {\bfseries\underline{always.}}  

\medskip

Besides, it is only the $S$ states of the charge that can contribute 
to the electron potential energy. The potential energy of an 
electron is mediated by the Coulomb interaction of the electron 
with the nucleus. We identify the nucleus with a geometric point 
placed at the origin of the spherical reference frame. The 
electron charge is distributed somehow in space. In this case, 
the potential energy of interaction of this charge with the 
nucleus can be written as
\begin{equation}
\label{1.1.4}
\mathcal E_{Pot}=\int e\delta(\mathbf{r})\Phi(\mathbf{r})dV,
\end{equation}
where $e\delta(\mathbf{r})$ is the nuclear charge, and  
$\Phi(\mathbf{r})$ is the potential 
generated by the electron charge distribution. Examining 
Eq. (\ref{1.1.4}), we see that the energy $\mathcal E_{Pot}$ 
is nonzero for the $S$ 
states only, because it is solely for the $S$ states that the 
potential $\Phi(0)$ is nonzero. While besides the $S$ states an atom 
may contain other charge states as well, they will not be able 
to contribute to the potential energy of electron interaction 
with the nucleus. This contribution can come from the $S$ states 
of charge only. 
 
\medskip 

Hence, {\bfseries\underline{in all situations}} the $S$ charge 
state  {\bfseries\underline{ is present}} in an atom.

The $S$ charge state {\bfseries\underline{ must be present}} in an 
atom {\bfseries\underline{always.}}  

\medskip

To find the charge shape assumed by an electron rotating about a 
nucleus, we write the corresponding equation in the form
\begin{equation}
\label{1.1.5}
\bigtriangleup \rho(\mathbf{r})+\frac{4m_e}{\hbar^2}
(\mathcal E-\mathcal E_{Pot})\rho(\mathbf{r})=0.
\end{equation}
In this equation, $\rho(\mathbf{r})$ is the density of the electron 
charge distribution we are looking for,  $m_e$ is the electron 
mass, $\mathcal E$ is the eigenvalue of energy, $\mathcal E_{Pot}$ 
is the electron potential energy, and $\hbar$ is the Dirac constant: 
$\hbar=h/2\pi$, where $h$ is the Planck constant. Equation 
(\ref{1.1.5}) differs in form from the Schr\"odinger equation 
in that it is written not for the wave function $\Psi$ but 
rather for the density of electron charge distribution 
$\rho(\mathbf{r})$, and in the numerical coefficient 4 in 
the second term of the equation (in the 
Schr\"odinger equation, the coefficient is 2). The meaning 
of this difference will become clear later. Equation (\ref{1.1.5}) 
may be called {\itshape\bfseries{ “the equation of wave mechanics”}}. 
This name stresses that an electron, while being point-like in free 
space and, hence, obeying the laws of mechanics, changes 
its shape in an atom as a result of manifestation of its 
wave properties when expanded in charge waves.

We are going to find the shape of the electron charge 
distribution by means of Eq. (\ref{1.1.5}). Equation (\ref{1.1.5}) 
contains the energy $\mathcal E_{Pot}$. This is the potential energy 
of an electron in the nuclear field. We do not know, however, the 
shape of the electron charge distribution and, hence, do not know 
the energy $\mathcal E_{Pot}$. The only thing we can do is to 
substitute for $\mathcal E_{Pot}$ the energy of interaction of a 
point nucleus with a point electron. In doing this, we have got 
ourselves into an ambiguity: indeed, we are looking for the shape 
of a distributed electron charge but substitute into Eq. (\ref{1.1.5}) 
the energy of a point electron. One should therefore verify 
that this will not give rise to an inconsistency. We shall 
check this later in the particular example of the hydrogen 
atom.

Introducing the parameter $\mu_e=2m_e$, Eq. (\ref{1.1.5}) can 
be recast in the form
\begin{equation}
\label{1.1.6}
\bigtriangleup \rho_\mu(\mathbf{r})+\frac{2\mu_e}{\hbar^2}
(\mathcal E_\mu-\mathcal E_{Pot})\rho_\mu(\mathbf{r})=0.
\end{equation}
This equation coinciding in form with the Schr\"odinger equation, 
one can use the results of its solution, keeping in mind all the 
time that Eq. (\ref{1.1.6}) contains in place of the mass $m_e$ the 
parameter $\mu_e$ which is not the electron mass. Therefore in 
what follows we are going to label all the results derived 
from Eq. (\ref{1.1.6}) with index $\mu$. The quantities labeled by 
$\mu$ may judiciously be called “modified” to stress that they are 
related not to the real electron mass $m_e$ but rather to a 
parameter $\mu_e$. Accordingly, Eq. (\ref{1.1.6}) may be referred 
to as a “modified” equation of wave mechanics.

It appears pertinent to specify now the relations connecting the 
commonly used quantities with the corresponding “modified” quantities.

\begin{center}Modified mass
$$
\mu_e=2m_e.
$$ 
Modified Bohr radius
\begin{equation}
\label{1.1.7}
a_\mu=\frac{\hbar^2}{\mu_ee^2}=\frac{\hbar^2}{2m_ee^2}=\frac{a}{2},
\end{equation} 
$$                                               
\mbox{where}\ a=\frac{\hbar^2}{m_ee^2} \mbox{ is the Bohr radius.} 
$$  
Modified reduced radius
$$
\tau_\mu=\frac{r}{a_\mu}=2\tau,
$$
\end{center} 
where $r$ is the radial coordinate of a spherical reference 
system, and $\tau=r/a$ is the reduced radius. We thus obtain 
$\tau_\mu a_\mu=\tau a=r$.

\section{Potential energy of distributed electron charge in 
a hydrogen atom}

Let us find the potential energy of interaction of an electron 
with the nucleus in a hydrogen atom. One should first find for 
this purpose the shape of the distributed charge of an electron 
in the potential field of the nucleus and the energy of interaction 
of this distributed charge with the nucleus by methods of 
electrodynamics. To calculate the charge distribution we are 
looking for, one has to substitute into the equation of wave 
mechanics (\ref{1.1.5}), or Eq. (\ref{1.1.6}), the potential energy 
of electron interaction with the nucleus. As pointed out in Sec. 1.2, 
the only possibility open for us here is to treat this energy as 
interaction of a point nucleus with a point electron. The potential 
energy of interaction of a point nucleus with a point electron 
can be written as
\begin{equation}
\label{1.2.1}
\mathcal E_{Pot}=-\frac{e^2}{r}
\end{equation}	
where $r$ is the distance between the nucleus and the electron. 
We assume the nucleus to be at the origin of the spherical 
coordinate system. In this case $r$ is nothing else but the radial 
component of the spherical reference frame. Substituting Eq. 
(\ref{1.2.1}) into Eq. (\ref{1.1.5}), we come to
\begin{equation}
\label{1.2.2}
\bigtriangleup \rho+\frac{4m_e}{\hbar^2}\left(\mathcal E+\frac{e^2}
{r}\right)\rho=0.
\end{equation}
One has now to solve this equation and find the charge distribution 
$\rho$. In place of solving Eq. (\ref{1.2.2}), however, we can write the 
“modified” equation by substituting Eq. (\ref{1.2.1}) into Eq. 
(\ref{1.1.6}), to obtain
\begin{equation}
\label{1.2.3}
\bigtriangleup \rho_\mu+\frac{2\mu_e}{\hbar^2}\left(\mathcal 
E_\mu+\frac{e^2}{r}\right)\rho_\mu=0.
\end{equation}
Recall that it is not the real electron mass $m_e$ but rather 
parameter $\mu_e=2m_e$ that enters this equation. Equation 
(\ref{1.2.3}) coincides in form with the Schr\"odinger equation 
for the hydrogen atom. The solutions to this equation are well 
known. In the form of most relevance to us they are presented in 
monograph \cite{2}. As pointed out in Sec. 1.2, of all the solutions 
we are interested in the spherical symmetric $S$ states only. 
We are writing out these solutions for the quantum numbers 
$n$ = 1, 2, 3 below (they are normalized against unity):
$$
\rho_{\mu1S}=2e^{-\tau_\mu},
$$
\begin{equation}
\label{1.2.4}
\rho_{\mu2S}=\frac{1}{\sqrt2}e^{-\frac{\tau_\mu}{2}}\left
(1-\frac{\tau_\mu}{2}\right),
\end{equation}
$$
\rho_{\mu3S}=\frac{2}{3\sqrt3}e^{-\frac{\tau_\mu}{3}}\left
(1-\frac{2}{3}\tau_\mu+\frac{2}{27}\tau_\mu^2\right).
$$
The variable in relations (\ref{1.2.4}) is the modified radius 
$\tau_\mu=r/a_\mu$, where $r$ is the radial component of the 
spherical coordinate system, and $a_\mu$ is the modified Bohr 
radius (see Eqs. (\ref{1.1.7})). Equations (\ref{1.1.7}) can 
be used to eliminate the $\mu$ index, with the solutions 
acquiring the form
$$
\rho_{1S}=A_{1S}e^{-2\tau},
$$
\begin{equation}
\label{1.2.5}
\rho_{2S}=A_{2S}e^{-\frac{2\tau}{2}}(1-\tau),
\end{equation}	  
$$
\rho_{3S}=A_{3S}e^{-\frac{2\tau}{3}}(27-36\tau+8\tau^2).
$$
We have introduced here normalization factors $A_{nS}$, 
so that there is no sense anymore in retaining the fractional 
coefficients in the parentheses. The variable in Eqs. (\ref{1.2.5}) 
is the reduced radius $\tau=r/a$. We are going to use besides 
$\tau$ in what follows the $r$ variable too. The solutions 
$\rho_{nS}$ are the densities of electron charge distribution 
in the atom. The total charge should be equal to the electron 
charge; therefore, the $\rho_{nS}$ functions should be normalized 
not to unity, as this is done in quantum mechanics, but rather 
to the electron charge $(-e)$:
\begin{equation}
\label{1.2.6}
\int \rho_{nS}dV=-e.
\end{equation}
$dV$ is here an element of volume in the spherical coordinate system:
\begin{equation}
\label{1.2.7}
dV=r^2\sin\vartheta d\vartheta d\varphi dr=a^2\tau^2\sin\vartheta 
d\vartheta d\varphi ad\tau=a^3\tau^2\sin\vartheta d\vartheta 
d\varphi d\tau.
\end{equation}
Integrating Eq. (\ref{1.2.6}), we come to the following values 
for the $A_{nS}$ coefficients:
$$
A_{1S}=\frac{-e}{\pi a^3},
$$
\begin{equation}
\label{1.2.8}
A_{2S}=\frac{+e}{4\pi a^3}\cdot \frac{1}{4},
\end{equation}
$$
A_{3S}=\frac{-e}{4\pi a^3}\cdot \frac{4}{27\cdot 27\cdot 3}.
$$
Having obtained the solutions (\ref{1.2.5}) for the electron 
charge distribution in an atom, we can find the energy 
of interaction of this charge with the nucleus:
\begin{equation}
\label{1.2.9}
\mathcal E_{Pot,nS}=\int \Phi_N\rho_{nS}dV.
\end{equation}
Here $\Phi_N=e/r=e/a\tau$ is the nuclear potential. 
Substituting now the expression for the nuclear 
potential and equations (\ref{1.2.5}) into Eq. (\ref{1.2.9}), and 
going over to a common variable $\tau$, we obtain after 
some straightforward calculations
\begin{equation}
\label{1.2.10}
\mathcal E_{Pot,1S}=-\frac{e^2}{a},\qquad \mathcal E_{Pot,2S}
=-\frac{e^2}{a}\cdot \frac{1}{4},\qquad \mathcal E_{Pot,3S}
=-\frac{e^2}{a}\cdot \frac{1}{9}.
\end{equation}
We have used in the integration the well known formula 
(valid for an integer $n$)
\begin{equation}
\label{1.2.11}
\int\limits_0^\infty x^ne^{-ax}dx=\frac{n!}{a^{n+1}}.
\end{equation}
The expressions for the energy (\ref{1.2.10}) coincide with the 
eigenvalues of Eq. (\ref{1.2.2}) although they were derived by 
another method.
The expressions (\ref{1.2.10}) can be combined in one formula
\begin{equation}
\label{1.2.12}
\mathcal E_{Pot,nS}=-\frac{e^2}{a}\cdot \frac{1}{n^2}.
\end{equation}
Equation (\ref{1.2.12}) coincides fully with the expression 
for the {\bfseries potential} energy of interaction between a nucleus 
and an electron which is well known in quantum mechanics 
(it could be obtained, for instance, from the expression 
$\mathcal E_{Pot,nS}=\int \Psi_{nS}^*\hat H \Psi_{nS} dV$, 
where $\hat H=-\frac{e^2}{r}$ is the potential energy operator).

Thus, by normalizing the charge distribution against the 
electron charge and using standard methods of electrostatics, 
we have arrived at correct values of the energies of interaction 
of a distributed charge with the nucleus, which coincide with 
those known from quantum mechanics.

\section{Fields and potentials of a distributed electron charge}

Knowing the distribution of electron charge in an atom, we can 
readily calculate the electric fields and potentials generated 
by these charges. As pointed out in Sec. 1.2, of all the solutions 
of the wave mechanics equation we are interested in spherically 
symmetric states of charge only. This simplifies greatly our task. 
Let us find now the potentials and fields of a distributed 
electron charge. The electric field can be derived from the Gauss’ 
theorem (see, e.g., Ref. \cite{3}, p. 109):
\begin{equation}
\label{1.3.1}
\oint\limits_S \mathbf E(\mathbf r)d\mathbf S=4\pi\int\limits_V\rho
(\mathbf r')dV'=4\pi Q',
\end{equation}
where for the surface $S$ we can take a sphere of radius 
$r$, $Q'$ is the charge inside this sphere, $\rho(\mathbf r')$ is 
the charge density, and $V$ is the volume inside the sphere $S$. 
Because by symmetry of the system the field on the surface of such 
a sphere is constant (and has only one radial component 
$E_r(r)$), the Gauss’ theorem takes on the form 
\begin{equation}
\label{1.3.2}
E(r)\cdot 4\pi r^2=4\pi\int\limits_V \rho(\mathbf r')dV',
\end{equation}
whence
\begin{equation}
\label{1.3.3}
E(r)=\frac{1}{r^2}\int\limits_V \rho(\mathbf r')dV'=\frac{4\pi}{r^2}
\int\limits_0^r \rho(r'){r'}^2dr'.
\end{equation}
The potential $\Phi(r)$ can be calculated from the formula
\begin{equation}
\label{1.3.4}
\Phi(r)=-\int\limits_\infty^r E(r')dr'.
\end{equation}
Substituting Eq. (\ref{1.3.3}) for the field $E(r)$ in this equality 
and integrating by parts, we come to the following expression 
for the potential
\begin{equation}
\label{1.3.5}
\Phi(r)=\frac{4\pi}{r}\int\limits_0^r \rho(r'){r'}^2dr'+4\pi\int
\limits_r^\infty \rho(r')r'dr'.
\end{equation}
Equations (\ref{1.3.3}) and (\ref{1.3.5}) can be found in monographs 
\cite{3} (pp. 113, 175) and \cite{4} (pp. 27, 228).

The expressions for distributed charge (\ref{1.2.5}) involve the 
variable $\tau=r/a$. Equation (\ref{1.3.5}) expressed in these 
variables assumes the form
\begin{equation}
\label{1.3.6}
\Phi(\tau)=\frac{4\pi a^2}{\tau}\int\limits_0^\tau \rho(\tau')
{\tau'}^2d\tau'+4\pi a^2\int\limits_\tau^\infty \rho(\tau')\tau'd\tau'.
\end{equation}
Substituting the expressions for distributed charge (\ref{1.2.5})  
in Eq. (\ref{1.3.6}), we come to the following relations for the 
potential of the distributed electron charge
$$
\Phi_{1S}=-\frac{e}{a\tau}+\frac{e}{a\tau}e^{-2\tau}(1+\tau),
$$
\begin{equation}
\label{1.3.7}
\Phi_{2S}=-\frac{e}{a\tau}+\frac{e}{a\tau}e^{-\frac{2\tau}{2}}
\cdot\frac{1}{4}(\tau^2+3\tau+4),
\end{equation}
$$
\Phi_{3S}=-\frac{e}{a\tau}+\frac{e}{a\tau}e^{-\frac{2\tau}{3}}
\cdot\frac{1}{27\cdot9}(8\tau^3+36\tau^2+27\cdot5\tau+27\cdot9).
$$
Recalling that $\tau=r/a$, these formulas can be made more 
revealing by a corresponding transformation
$$
\Phi_{1S}=-\frac{e}{r}+\frac{e}{r}e^{-2\tau}(1+\tau),
$$
\begin{equation}
\label{1.3.8}
\Phi_{2S}=-\frac{e}{r}+\frac{e}{r}e^{-\frac{2\tau}{2}}\cdot
\frac{1}{4}(\tau^2+3\tau+4),
\end{equation}
$$
\Phi_{3S}=-\frac{e}{r}+\frac{e}{r}e^{-\frac{2\tau}{3}}\cdot
\frac{1}{27\cdot9}(8\tau^3+36\tau^2+27\cdot5\tau+27\cdot9).
$$
Examining expressions (\ref{1.3.8}), we see immediately that 
the potentials of distributed electron charge in a hydrogen 
atom are actually a sum of the potential of a point charge 
$(-e)$ placed at the origin and the potential of a point charge 
$(+e)$ but with an exponential factor, likewise located at the origin.

Consider now the limiting cases of the behavior of the potential. 
For $r\to\infty$, the second term in expressions (\ref{1.3.8}) vanishes by virtue 
of the exponential factor, with the potentials reducing to a 
potential of a point charge $(-e)$ placed at the origin. For $r\to0$,
 the exponentials can be expanded in a series, with two terms 
 retained. This leads us to
\begin{equation}
\label{1.3.9}
\Phi_{1S}(0)=-\frac{e}{a},\qquad \Phi_{2S}(0)=-\frac{e}{a}\cdot
\frac{1}{4},\qquad \Phi_{3S}(0)=-\frac{e}{a}\cdot\frac{1}{9}.
\end{equation}
Because we assume the nucleus to be point-like, located at the 
origin, and having a charge $q_N=(+e)$, the potential energy 
of interaction of the nucleus with a distributed electron 
charge can be derived without integration
\begin{equation}
\label{1.3.10}
\mathcal E_{nS}=q_{N}\Phi_{nS}(0)=e\Phi_{nS}(0).
\end{equation}
This yields for the interaction energy
\begin{equation}
\label{1.3.11}
\mathcal E_{1S}=-\frac{e^2}{a},\qquad \mathcal E_{2S}=-\frac{e^2}
{a}\cdot\frac{1}{4},\qquad \mathcal E_{3S}=-\frac{e^2}{a}\cdot
\frac{1}{9}.
\end{equation}
These expressions coincide naturally with Eqs. (\ref{1.2.10})
 calculated from Eq. (\ref{1.2.9}).
 
Calculate now the fields corresponding to the charge distributions 
obtained. By virtue of the spherical symmetry of the charge 
distribution, the fields will have only one radial component 
$E_r$: $E_r=-\bigtriangledown_r\Phi=-\frac{1}{a}
 \bigtriangledown_\tau\Phi$. Taking a derivative of expressions 
(\ref{1.3.7}), we come to 
$$
E_{1S}=-\frac{e}{a^2\tau^2}+\frac{e}{a^2\tau^2}e^{-2\tau}
(2\tau^2+2\tau+1),
$$
\begin{equation}
\label{1.3.12}
E_{2S}=-\frac{e}{a^2\tau^2}+\frac{e}{a^2\tau^2}e^{-\tau}
\cdot \frac{1}{4}(\tau^3+2\tau^2+4\tau+4),
\end{equation}
$$
E_{3S}=-\frac{e}{a^2\tau^2}+\frac{e}{a^2\tau^2}e^{-\frac{2\tau}{3}}
\cdot\frac{1}{27\cdot27}(16\tau^4+24\tau^3+9\cdot18\tau^2+27
\cdot18\tau+27\cdot27).
$$
Just as in the case with the potential, we rewrite Eqs. 
(\ref{1.3.12}) in a more revealing way
$$
E_{1S}=-\frac{e}{r^2}+\frac{e}{r^2}e^{-2\tau}(2\tau^2+2\tau+1),
$$
\begin{equation}
\label{1.3.13}
E_{2S}=-\frac{e}{r^2}+\frac{e}{r^2}e^{-\tau}\cdot \frac{1}{4}
(\tau^3+2\tau^2+4\tau+4),
\end{equation}
$$
E_{3S}=-\frac{e}{r^2}+\frac{e}{r^2}e^{-\frac{2\tau}{3}}\cdot
\frac{1}{27\cdot27}(16\tau^4+24\tau^3+9\cdot18\tau^2+27
\cdot18\tau+27\cdot27).
$$
Examining Eqs. (\ref{1.3.13}), we see that in all cases the 
field of a distributed electron charge may be
considered as a sum of two fields, more specifically, 
of a field of a point charge $(-e)$ located at the
origin plus that of a point charge $(+e)$ but with an 
exponential factor, which is likewise placed at the
origin.

Consider now the extreme cases. For $r\to\infty$, the second 
term in all expressions vanishes by virtue of the exponential 
factor, to leave the field of a point charge $(-e)$ at the 
origin. To study the behavior of the field for $r\to0$, one 
can expand the exponential to third order. Substituting this 
expansion in Eqs. (\ref{1.3.12}), we readily see that all fields 
in this case vanish. 

The expressions for the potential $\Phi_{1S}$ and field $E_{1S}$ of 
the $1S$ state can be found in Refs. \cite{3} (pp. 113, 176) 
and \cite{4} (pp. 27 and 228). 

The field acting in the atom is actually a sum of the field 
of the nucleus $E_N=e/r^2$ and of that created by the distributed 
charge (\ref{1.3.12}), (\ref{1.3.13}). At a certain distance from 
the nucleus this total field will tend to zero by virtue of the 
exponential factor. As one approaches the nucleus, the total field 
resembles the field of the nucleus $e/r^2$.

Similarly, the potential acting in the atom is a sum of that 
created by the nucleus $\Phi_N=e/r$ and of the distributed charge 
potential (\ref{1.3.7}), (\ref{1.3.8}). At some distance from the 
nucleus, this total potential tends to zero by virtue of the 
exponential factor. As one comes closer to the nucleus, the total 
potential approaches in form the nuclear potential $e/r$.

\section{Kinetic and total energies of a distributed electron charge 
in an atom} 

We have seen that the charge which was point-like in free space 
spreads out in an atom to become a distributed rather that point 
charge. In free space, however, electron has not only a point charge 
but a point mass as well. On entering an atom, this point mass has 
to spread out just as this was done by the point charge. The shape 
of the mass distribution should be analogous to that of the charge 
distribution, because it forms by summation of absolutely identical 
harmonics both in the first and the second cases. Indeed, a point 
electron possesses both a point charge and a point mass. The point 
charge and mass can be expanded in terms of harmonics as this was 
done in series (\ref{1.1.1}) or (\ref{1.1.3}); a difference may 
appear only in the common coefficient and the common sign of the 
harmonics. The velocities and directions of motion of the harmonics 
also coincide, because both the charge and the mass belong to the 
same point electron. Therefore, as the electron moves in its circular 
trajectory, summation of all harmonics in which the mass was expanded 
should produce the same distribution of mass as that of the charge 
(to within the common coefficient and sign). In other words, we deal 
here not with a distribution of charge or that of mass but rather 
with a distribution of {\itshape\bfseries the charge/mass object}. 
Therefore, we can write an equation for the distribution of mass 
similar to Eq. (\ref{1.1.5}):
\begin{equation}
\label{1.4.1}
\bigtriangleup m(\mathbf{r})+\frac{4m_e}{\hbar^2}(\mathcal 
E-\mathcal E_{Pot})m(\mathbf{r})=0.
\end{equation}
For the hydrogen atom, this equation can be recast to the form 
of Eq. (\ref{1.2.2}):
$$
\bigtriangleup m+\frac{4m_e}{\hbar^2}\left(\mathcal E+
\frac{e^2}{r}\right)m=0.
$$
We write the solutions to this equation in a form similar to that 
of Eqs. (\ref{1.2.5}):
$$
m_{1S}=B_{1S}e^{-2\tau},
$$
\begin{equation}
\label{1.4.2}
m_{2S}=B_{2S}e^{-\frac{2\tau}{2}}(1-\tau),
\end{equation}	  
$$
m_{3S}=B_{3S}e^{-\frac{2\tau}{3}}(27-36\tau+8\tau^2).
$$
To find the coefficients $B_{nS}$, one has first to normalize 
the $m_{nS}$ functions against the electron mass $m_e$:
\begin{equation}
\label{1.4.3}
\int m_{nS}dV=m_e.
\end{equation}
Integration yields the following expressions for the $B_{nS}$ 
coefficients
$$
B_{1S}=\frac{m_e}{\pi a^3},
$$
\begin{equation}
\label{1.4.4}
B_{2S}=\frac{-m_e}{4\pi a^3}\cdot \frac{1}{4},
\end{equation}
$$
B_{3S}=\frac{m_e}{4\pi a^3}\cdot \frac{4}{27\cdot 27\cdot 3}.
$$
Thus, we have found the shape of the mass distribution. 
To be precise, this is not the shape of the mass distribution 
but rather that of {\itshape\bfseries the charge/mass object}. 
What’s more, examining Eqs. (\ref{1.2.8}) and (\ref{1.4.4}) we 
see that the distributions of charge and mass are related through
\begin{equation}
\label{1.4.5}
m(\mathbf r)=-\frac{m_e}{e} \rho(\mathbf r).
\end{equation}
In what follows, in all cases where the term “charge” or “mass” 
appears it will be assumed that in actual fact this is 
{\itshape\bfseries the charge/mass object}.

A comment is appropriate here. The mass distributions (\ref{1.4.2}) 
are normalized against the electron mass, i.e., they are always 
positive. Equations (\ref{1.4.2}) contain, however, alternating 
polynomials, with the result that within some intervals the mass 
may acquire a negative sign. There is nothing particular in this 
from the mathematical side of view, but theoretical mechanics does 
not operate with such a notion as negative mass. Most probably, 
the “negativeness” of the mass may physically come up in combination 
with some other parameters, for instance, in the momentum or angular 
momentum. The momentum and angular momentum may have either sign, 
and the negative sign means only motion or rotation in the opposite 
direction.

Besides the shape of the mass distribution, one can envisage some 
other parameters as well which are connected with mass. Indeed, 
electron in an atom possesses kinetic energy. Hence, its mass should 
move somehow. The only internal motion allowed for an atom is rotation. 
In other words, we have to find the form of this rotation. But 
neither Eq. (\ref{1.4.1}) nor Eq. (\ref{1.1.5}) can yield the form 
of rotation, because these relations do not contain any parameters 
of motion at all. But some form of rotation has to be present in an 
atom. There is, however, another consideration.

By the theorem of Earnshaw, a stable static configuration of electric 
charges cannot exist without involvement of any other forces of 
other than electric origin. For this reason, static charge 
distributions (\ref{1.2.5}) are intrinsically unstable. But the atom 
is stable. Only rotation can impart stability to the atom. Hence, 
the charges described by Eqs. (\ref{1.2.5}) should rotate.

Motion (considered in nonrelativistic approximation) cannot change 
the potential energy of a distributed charge. The potential energy 
of a rotating charge distribution can be readily found assuming the 
charge being at rest. Indeed, any charge which has departed from a 
given point is replaced immediately by an identical charge at this 
point. Potential energy depends only on the magnitude of the charge 
at a given point and is independent of whether this is the charge 
that has just left or the second (that has arrived).
\medskip

Let us look now for the overall pattern of rotation of a distributed 
charge and a distributed mass. What are the basic considerations 
we should start from? First, we know from quantum mechanics the 
expression for the kinetic energy of a rotating electron 
$\mathcal E_{Kin}$ (derived, for instance, from the relation 
$\mathcal E_{Kin,nS}=\int\Psi_{nS}^* \hat H \Psi_{nS} dV$,
 where $\hat H=\frac{\hat p^2}{2m}$ is the kinetic energy operator, 
 and $\hat p=-i\hbar\bigtriangledown$ is the momentum operator): 
\begin{equation}
\label{1.4.6}
\mathcal E_{Kin,nS}=\frac{e^2}{2a}\cdot\frac{1}{n^2}.
\end{equation}
Second, quantum mechanics offers the following expression for the 
angular momentum of an electron in the $S$ states, $M_{nS}$:
\begin{equation}
\label{1.4.7}
M_{nS}=const=\hbar/2,
\end{equation}
because an electron has in the $S$ states only the spin moment 
$\hbar/2$. (Recall that nonrelativistic quantum mechanics required 
that the angular momenta of $S$ states be zero, while experiment 
showed them to be $\hbar/2$ rather than zero.) Thus, rotation should 
be such as to satisfy at least these two conditions, 
(\ref{1.4.6}) and (\ref{1.4.7}).

The simplest assumption that comes immediately to mind is as follows. 
The distributions of mass and charge rotate as a whole, i.e., as 
a continuous body. Let us assume that rotation occurs around a vertical 
axis $(\vartheta=0)$. Let us call it for convenience the $Z$ axis, and 
the plane $\vartheta=\pi/2$, the equatorial plane. In this case, the 
velocity $v_\varphi$ of motion of any point is proportional to the 
radius and the sine of the angle $\vartheta$:
\begin{equation}
\label{1.4.8}
v_\varphi=Ar\sin\vartheta=Aa\tau\sin\vartheta,
\end{equation}
where $A$ is a constant. The above suggests that as one moves away 
from the axis of rotation, $v_\varphi$ {\bfseries increases}, and as 
one approaches this axis, it {\bfseries decreases}. 

Let us verify that this assumption meets conditions (\ref{1.4.6}) 
and (\ref{1.4.7}). Substituting the expressions for the mass (\ref{1.4.2}) 
in the relation
\begin{equation}
\label{1.4.9}
\mathcal E_{Kin,nS}=\int\frac{m_{nS}v_\varphi^2}{2}dV,
\end{equation}
where the element of volume $dV$ is defined by Eq. (\ref{1.2.7})
we come to
\begin{equation}
\label{1.4.10}
\mathcal E_{Kin,1S}=m_eA^2a^2\cdot1,\quad \mathcal E_{Kin,2S}
=m_eA^2a^2\cdot8,\quad \mathcal E_{Kin,3S}=m_eA^2a^2\cdot57.
\end{equation}
We calculate now the angular momentum around the $Z$ axis
$(\vartheta=0)$
\begin{equation}
\label{1.4.11}
M_{nS}=\int m_{nS}v_\varphi r\sin\vartheta dV.
\end{equation}
For the momenta $M_{nS}$ we obtain
\begin{equation}
\label{1.4.12}
M_{1S}=m_eAa^2\cdot2,\qquad M_{2S}=m_eAa^2\cdot16,\qquad 
M_{3S}=m_eAa^2\cdot114.
\end{equation}
Examining now Eqs. (\ref{1.4.10}) and (\ref{1.4.12}), we see that 
the energies $\mathcal E_{Kin,nS}$ do not scale as $1/n^2$, and that 
the angular momenta $M_{nS}$ are in no way constant. 
Said otherwise, the conditions (\ref{1.4.6}) and (\ref{1.4.7}) are 
not met. But this means that the distributions of charge 
and mass {\bfseries \underline{cannot rotate as a whole}}, 
i.e., as a solid  body.
\medskip

To establish the pattern of rotation of the charge and mass 
that can exist in an atom, the following reasoning appears to 
be appropriate. Consider circular rotation of an element of 
mass $dm$ with a negative charge $dq$ in the equatorial plane about 
a nucleus with charge $(+e)$ as about the center.

Rotation of an element of mass $dm$ along a circle of radius $R$ 
is driven by the action on the element of mass of a centripetal 
force $d\mathbf F$ toward the center: 
$d\mathbf F=-\frac{dmv_\varphi^2}{R}\mathbf n$, where $v_\varphi$ 
is the velocity of the element of mass, and $\mathbf n$ is the unit 
vector in the direction of the radius $R$. The negative sign of 
the centripetal force  $d\mathbf F$ signifies that the force is 
directed oppositely to the unit vector $\mathbf n$. In our 
particular case, the centripetal force is essentially the force 
of attraction between the charges, $d\mathbf F=\frac{edq}{R^2}\mathbf n$. 
Equating these two expressions, canceling $R$ and dividing by two, 
we finally obtain 
\begin{equation}
\label{1.4.13}
\frac{dmv_\varphi^2}{2}=\frac{1}{2}\cdot\left(-\frac{edq}{R}\right).
\end{equation}
We see on the left the kinetic energy of the element of mass, 
and on the right, one half of the potential energy taken with 
the opposite sign. This means that circular rotation of an 
element of mass $dm$ obeys the relation
\begin{equation}
\label{1.4.14}
\mathcal E_{Kin}=\frac{1}{2}(-\mathcal E_{Pot}).
\end{equation}
The equality (\ref{1.4.14}) formulates the Clausius theorem on 
the virial of forces mediating circular motion of an element 
of mass in a Coulomb potential well (see, e.g., \cite{5}, p. 76). 
Because the potential, kinetic, and total energies of an element 
of mass moving circularly are constant, there is
no need in averaging the energies, as this would be required 
by the theorem of virial in a general form.

Let us write now the theorem of virial for an element or sum of 
elements of mass/charge moving in a Coulomb potential well in a 
general form (see, e.g., \cite{5}):
\begin{equation}
\label{1.4.15}
\overline{\mathcal E_{Kin}}=\frac{1}{2}(-\overline{\mathcal 
E_{Pot}}),
\end{equation}
where the bar on top signifies averaging over time.

Equation (\ref{1.4.13}) yields the velocity of circular motion 
of an element $dm$ (at the equator)
\begin{equation}
\label{1.4.16}
v_\varphi=\sqrt{-\frac{e}{R}\cdot\frac{dq}{dm}},
\end{equation}
and, combined with Eq. (\ref{1.4.5}), we finally have
\begin{equation}
\label{1.4.17}
v_\varphi=\frac{e}{\sqrt{m_e}}\cdot\frac{1}{\sqrt{R}},
\end{equation}
i.e., the velocity of an element of mass is inversely 
proportional to the square root of the distance between 
the nucleus and the element. Thus, as one approaches the 
axis of rotation, {\bfseries\underline{the velocity increases}}, 
and as one moves away from it, 
{\bfseries\underline{the velocity decreases}}.

Because at the equator the radius $R$ of the circle along 
which the element of mass moves coincides with the coordinate 
$r$ of the spherical reference frame, Eq. (\ref{1.4.17}) can be 
recast in the form
\begin{equation}
\label{1.4.18}
v_\varphi=\frac{\alpha c}{\sqrt{\tau}}.
\end{equation}
In this expression, the radius of the circle $R=r=a\tau$, where 
$a$ is, as before, the Bohr radius: $a=\frac{\hbar^2}{m_ee^2}$, 
$\alpha$ is the fine structure constant: $\alpha=\frac{e^2}{\hbar c}$, 
and $c$ is the velocity of light.

Finding the velocity of an element of mass not lying in the 
equatorial plane meets with some difficulties. In this case, 
the centripetal force does not coincide in direction with the 
force of attraction to the nucleus, and this makes the above 
reasoning invalid here.

To describe the rotation of a distributed charge as a whole, 
we can make two assumptions: 1. Rotation occurs in such a way 
that linear velocity $v_\varphi$ of each element depends only 
on its distance from the nucleus; 2. Rotation occurs such that 
elements located on the sphere of radius $r$ have the same 
angular velocity. In the first case, the relation for the 
velocity of motion of elements of a distributed charge can be 
written in the way similar to Eq. (\ref{1.4.18})
\begin{equation}
\label{1.4.19}
v_\varphi=\frac{k'\alpha c}{\sqrt{\tau}}.
\end{equation}
In the second case, the equation assumes the form
\begin{equation}
\label{1.4.20}
v_\varphi=\frac{k''\alpha c}{\sqrt{\tau}}\sin\vartheta.
\end{equation}
Here $k'$ and $k''$ are some coefficients. In both cases, 
{\itshape\bfseries the charge rotates in a layered pattern 
depending on the radius} $r$, with {\itshape\bfseries the 
velocity of rotation being the higher, the closer is the 
charge element to the nucleus}. 

Thus, of the two versions of the velocity, (\ref{1.4.19}) 
and (\ref{1.4.20}), we have to choose the right one.

Substituting the velocity from Eq. (\ref{1.4.19}) in Eq. 
(\ref{1.4.9}), we obtain
\begin{equation}
\label{1.4.21}
\mathcal E_{Kin,1S}=k'^2\frac{e^2}{2a}\cdot1,\quad \mathcal 
E_{Kin,2S}=k'^2\frac{e^2}{2a}\cdot\frac{1}{4},\quad \mathcal 
E_{Kin,3S}=k'^2\frac{e^2}{2a}\cdot\frac{1}{9}.
\end{equation}

Substituting now the velocity from Eq. (\ref{1.4.20}) in Eq. 
(\ref{1.4.9}), we come to
\begin{equation}
\label{1.4.22}
\mathcal E_{Kin,1S}=k''^2\frac{e^2}{2a}\cdot\frac{2}{3}\cdot1,
\quad \mathcal E_{Kin,2S}=k''^2\frac{e^2}{2a}\cdot\frac{2}{3}
\cdot\frac{1}{4},\quad \mathcal E_{Kin,3S}=k''^2\frac{e^2}{2a}
\cdot\frac{2}{3}\cdot\frac{1}{9}.
\end{equation}
An analysis of Eqs. (\ref{1.4.21}) and (\ref{1.4.22}) suggests 
that the energies $\mathcal E_{Kin,nS}$ scale as $1/n^2$, as 
required by Eq. (\ref{1.4.6}).

The virial theorem (\ref{1.4.15}) permits us now to find the 
$k'$ and $k''$ coefficients. We come eventually to $k'=1$,  
$k''=\sqrt{3/2}$.

For the kinetic energy we obtain in the two cases the following 
expressions
\begin{equation}
\label{1.4.23}
\mathcal E_{Kin,1S}=\frac{e^2}{2a}\cdot1,\qquad \mathcal E_{Kin,2S}
=\frac{e^2}{2a}\cdot\frac{1}{4},\qquad \mathcal E_{Kin,3S}
=\frac{e^2}{2a}\cdot\frac{1}{9}.
\end{equation}
Equations (\ref{1.4.23}) can be combined to yield 
\begin{equation}
\label{1.4.24}
\mathcal E_{Kin,nS}=\frac{e^2}{2a}\cdot\frac{1}{n^2}.
\end{equation}
Equation (\ref{1.4.24}) coincides exactly with Eq. (\ref{1.4.6}) 
for the kinetic energy of an electron, which is well known from 
quantum mechanics.

Let us determine now the total energy of the electron. By adding 
the potential energy (\ref{1.2.10}) just found with the kinetic 
energy (\ref{1.4.23}), we arrive at the total electron energy
\begin{equation}
\label{1.4.25}
\mathcal E_{1S}=-\frac{e^2}{2a}\cdot1,\qquad \mathcal 
E_{2S}=-\frac{e^2}{2a}\cdot\frac{1}{4},\qquad \mathcal 
E_{3S}=-\frac{e^2}{2a}\cdot\frac{1}{9}.
\end{equation}
Expressions (\ref{1.4.25}) can now be combined 
\begin{equation}
\label{1.4.26}
\mathcal E_{nS}=-\frac{e^2}{2a}\cdot\frac{1}{n^2},
\end{equation}
in an expression for the total electron energy, likewise 
well known from quantum mechanics.

Thus, basing on the formulas and methods of electrodynamics 
and mechanics and applying the above approach, we have come 
to absolutely correct values of the potential, kinetic, and 
total energies of an electron in a hydrogen atom. Significantly, 
in so doing we have not invoked such purely quantum-mechanical 
concepts as the wave function and the operator.
\medskip

Let us calculate now the angular momentum of a distributed 
mass about a vertical axis $Z$ $(\vartheta=0)$.

Substituting the velocity $v_\varphi$ from Eqs. (\ref{1.4.19}) 
and (\ref{1.4.20}) into the expression for the angular momentum 
(\ref{1.4.11}) we come, respectively, to
\begin{equation}
\label{1.4.27}
M_{1S}=0.92\hbar,\qquad M_{2S}=1.63\hbar,\qquad M_{3S}=2.40\hbar,
\end{equation}
and
\begin{equation}
\label{1.4.28}
M_{1S}=0.96\hbar,\qquad M_{2S}=1.70\hbar,\qquad M_{3S}=2.49\hbar.
\end{equation}
To calculate the radial integrals, we have to use now, in place of 
Eq. (\ref{1.2.11}) valid for an integer exponent $n$, a more 
general formula
\begin{equation}
\label{1.4.29}
\int\limits_0^\infty e^{-ax}x^ndx=\frac{\Gamma(n+1)}{a^{n+1}}
\qquad (\mbox{при }a>0 \quad n>-1),
\end{equation}
because the exponent of $x$ assumes not integer but rather 
half-integer values. $\Gamma(n+1)$ is the gamma function.

Examining now Eqs. (\ref{1.4.23}) and (\ref{1.4.24}), we see 
that we have obtained correct values for the kinetic energy 
of the electron and a correct dependence on number $n$. As for 
the angular momenta, they are not equal to $\hbar/2$, and, 
more than that, they are not equal to a constant value at all.

Hence, the assumptions concerning the velocity of rotation of a 
distributed charge are not accurate, and this problem requires 
a further study.

In Chapter 2 of this Paper, the velocity of rotation of a 
distributed charge is analyzed in more detail, and it is 
demonstrated that the angular momentum of the ground state 
is $\hbar/2$, in full agreement with experimental observations.

%% file: ch2.tex
\chapter{Non-spherical atom}
\label{ch2}

\section{Shape of a non-spherical charge}

As shown in Chapter 1 of the present Paper, if the electron 
charge distribution is assumed to be spherically symmetric, 
one cannot obtain a correct value of the projection of angular 
momentum on the $Z$ axis. One may therefore suggest that the 
distribution of the charge/mass does depend somehow on the 
angle $\vartheta$. We have not as yet, however, an equation 
more accurate than Eq. (\ref{1.1.5}) of wave mechanics and, 
therefore, we cannot know a more accurate solution. This 
implies that the assumption concerning the actual form of 
the dependence of the distribution on the angle  $\vartheta$ 
will have to be chosen intuitively. This also means that we 
can no longer use Eq. (\ref{1.2.2}). We are going to employ, 
however, the main conclusions derived by means of this equation. 
We shall apply, in particular, the radial dependence of charge 
density which was derived for a spherically symmetric charge 
distribution.

Consider the hydrogen atom in ground state. 

The simplest assumptions that appear reasonable in this case 
consist in that the distributed charge scales with $\vartheta$ 
as $\sin\vartheta$ or $\sin^2\vartheta$. In these conditions, 
the charge density can be written as follows
\begin{equation}
\label{2.1.1}
\rho'_{1NS}=A'_{1NS}\cdot e^{-2\tau}\cdot\sin\vartheta,
\end{equation}
or
\begin{equation}
\label{2.1.2}
\rho''_{1NS}=A''_{1NS}\cdot e^{-2\tau}\cdot\sin^2\vartheta.
\end{equation}
Here the subscript $1NS$ identifies the state which is similar 
to $1S$ but not spherically symmetric.

The coefficients $A'_{1NS}$ and $A''_{1NS}$ are found by 
normalization against the electron charge $(-e)$
\begin{equation}
\label{2.1.3}
\int \rho'_{1NS}dV=-e,\qquad \int \rho''_{1NS}dV=-e.
\end{equation} 
We finally arrive at
\begin{equation}
\label{2.1.4}
A'_{1NS}=\frac{-4e}{a^3\pi^2},\qquad A''_{1NS}=\frac{-3e}{2\pi a^3}.
\end{equation}

One may as reasonably assume that the angular dependence can be 
a combination of several spherical functions $Y_{lm}$. As shown in 
Chapter 1, the $l=0$ spherical function must be present in an atom 
always, because it is only this function that permits description 
of the charge in an atom. All the other functions are capable of 
affecting the charge shape only, without adding or subtracting any 
charge.

What other functions could be used to describe a distributed charge 
in an atom? These functions should be symmetric with respect to the 
angle $\vartheta=\pi/2$; indeed, there are no grounds to assume that 
an atom can be asymmetric relative to the equator, unless some 
additional external fields interfere. Besides, we can select for 
description of the charge only functions with $m=0$. Functions with 
$m\ne0$ have no axial symmetry relative to the $Z$ axis; therefore, 
any rotation of such an asymmetric charge should give rise to 
emission of radiation, which in actual fact does not happen.

The simplest function satisfying these requirements is $Y_{20}$. 
Therefore, we write the angular part of the formula for the 
charge distribution in the following form:
\begin{equation}
\label{2.1.5}
L=D(Y_{00}+D_{20}Y_{20}),
\end{equation}
$$
\mbox{where}\ Y_{00}=\frac{1}{\sqrt{4\pi}},\ Y_{20}
=\sqrt{\frac{5}{4\pi}}\left(\frac{3}{2}\cos^2\vartheta-\frac{1}
{2}\right),\ \mbox{(see, for instance, Ref. \cite{2}).}
$$
Coefficient $D_{20}$ can be found from the condition that 
function $L$ for $\vartheta=0$ and $\vartheta=\pi$ (i.e., at the 
$Z$ axis) be zero. Indeed, if a charge rotates about the $Z$ axis, 
all elements of charge not on the $Z$ axis are acted upon by 
both attraction to the nucleus and the centrifugal force. On 
the $Z$ axis, the centrifugal force is zero, therefore the 
only possibility for existence of a distributed charge lies 
in the absence of charge on the $Z$ axis.

Coefficient $D$ will be found from the condition that function 
$L$ for $\vartheta=\pi/2$ (i.e., at the equator) be unity. 
We obtain from these conditions
$$
D=\frac{2\sqrt{4\pi}}{3},\qquad D_{20}=-\frac{1}{\sqrt{5}}.
$$
Thus, function $L$ acquires the form
$$
L=\frac{2\sqrt{4\pi}}{3}\left[\frac{1}{\sqrt{4\pi}}-\frac{1}
{\sqrt5}\cdot\sqrt{\frac{5}{4\pi}}\left(\frac{3}{2}\cos^2
\vartheta-\frac{1}{2}\right)\right].
$$
One can easily verify that this function exactly coincides with 
the function $\sin^2\vartheta$. In other words, we obtain in 
this case two forms for the density of distributed charge:
\begin{equation}
\label{2.1.6}
\rho''_{1NS}=A''_{1NS}e^{-2\tau}\sin^2\vartheta,
\end{equation}
or
\begin{equation}
\label{2.1.7}
\rho''_{1NS}=A''_{1NS}e^{-2\tau}D(Y_{00}+D_{20}Y_{20}).
\end{equation}
One can chose conveniently the form most suitable for the 
actual conditions.

Let us see whether we obtain a correct potential energy of 
interaction of the nucleus with these forms of distributed 
charge. The potential produced by a distributed charge at 
the nucleus, i.e., at the origin, can be written as 
\begin{equation}
\label{2.1.8}
\Phi'(0)=\int\frac{\rho'_{1NS}}{r}dV=\int\frac{\rho'_{1NS}}{a\tau}dV,
\quad \Phi''(0)=\int\frac{\rho''_{1NS}}{r}dV=\int\frac{\rho''_{1NS}}
{a\tau}dV.
\end{equation}
We consider the nucleus to be a point at the origin, with the 
charge $(+e)$. Therefore, for the energy of interaction of 
the nucleus with the distributed electron charge can be written as
\begin{equation}
\label{2.1.9}
\mathcal E'_{Pot}=e\Phi'(0),\qquad \mathcal E''_{Pot}=e\Phi''(0).
\end{equation}
One can easily verify that in both cases we arrive at the same 
result:
\begin{equation}
\label{2.1.10}
\mathcal E'_{Pot}=\mathcal E''_{Pot}=-\frac{e^2}{a},
\end{equation}
which coincides with the energy of the spherically symmetric 
$1S$ state (\ref{1.2.10}), as well as with the results known 
from quantum mechanics.
\medskip

Let us calculate now the kinetic energy of a rotating charge. 
On repeating the arguments concerning the mass distribution 
formulated for a spherically symmetric charge (Sec. 1.5), 
we come to the conclusion that the mass distribution 
corresponding to the distribution of charge (\ref{2.1.1}) 
has the form
\begin{equation}
\label{2.1.11}
m'_{1NS}=B'_{1NS}e^{-2\tau}\sin\vartheta,
\end{equation}
and that the distribution of mass corresponding to the charge 
distribution (\ref{2.1.6}) or (\ref{2.1.7}) reads as 
\begin{equation}
\label{2.1.12}
m''_{1NS}=B''_{1NS}e^{-2\tau}\sin^2\vartheta,
\end{equation}
or
\begin{equation}
\label{2.1.13}
m''_{1NS}=B''_{1NS}e^{-2\tau}D(Y_{00}+D_{20}Y_{20}).
\end{equation}
The coefficients $B'_{1NS}$ and $B''_{1NS}$ are found by 
normalizing the mass density by the electron mass $m_e$. 
In this way we obtain for the $B'_{1NS}$ and $B''_{1NS}$ 
coefficients
\begin{equation}
\label{2.1.14}
B'_{1NS}=\frac{4m_e}{a^3\pi^2},\qquad B''_{1NS}=\frac
{3m_e}{2\pi a^3}.
\end{equation}
This distribution rotates about the $Z$ axis.

Let us find now the rotation velocity $v_\varphi$.

Two versions of the rotation velocity, (\ref{1.4.19}) and 
(\ref{1.4.20}), were proposed for the spherical charge 
distribution. If a charge distribution has no spherical 
symmetry (Eqs. (\ref{2.1.1}) or (\ref{2.1.2})), there is no 
charge near the $Z$ axis. This means that we may remove 
now the assumption of the angular velocity for charges 
on a sphere of radius $r$ being constant (as this was done 
in Chapter 1 of the Paper) and retain only the dependence 
of the velocity on radius. The dependence of the velocity 
on radius only can be assigned to the fact that the only 
source of the force propelling the motion of a distributed 
charge is the nucleus. Therefore (in contrast to Eq. 
(\ref{1.4.20})), we write the expression for the velocity 
of a distributed charge, similar to Eq. (\ref{1.4.19}), 
for $k'=1$ in the following form
\begin{equation}
\label{2.1.15}
v_\varphi=\frac{\alpha c}{\sqrt\tau},
\end{equation}
where  $\alpha$ is the fine structure constant, $c$ is the 
velocity of light, and $\tau=r/a$ is the reduced radius.

Equation (\ref{2.1.15}) shows that the velocity $v_\varphi$ 
increases as one approaches the nucleus. The charge rotates, 
as before, in a stratified fashion, but now it is the linear 
velocity $v_\varphi$ along the $\varphi$ coordinate rather 
than the angular velocity that depends on radius. We meet, 
however, as before, with a difficulty of identifying the 
factor that causes the motion of charges along the $\varphi$ 
coordinate if the force center does not lie in the $\varphi$ 
orbital plane (except for the equator). We still have not got, 
however, any other pattern of rotation to compare.

Calculate now the kinetic energy of a rotating charge. 
The kinetic energy  $\mathcal E_{Kin}$ can be written as   
\begin{equation}
\label{2.1.16}
\mathcal E_{Kin}=\int\frac{mv^2_\varphi}{2}dV.
\end{equation}
Substituting Eqs. (\ref{2.1.11}), (\ref{2.1.12}), and 
(\ref{2.1.15}) in Eq. (\ref{2.1.16}), we come to
\begin{equation}
\label{2.1.17}
\mathcal E'_{Kin}=\mathcal E''_{Kin}=\frac{1}{2}\cdot\frac{e^2}{a},
\end{equation}
which coincides with the formula derived for a spherically 
symmetric $1$S charge/mass distribution (\ref{1.4.23}) and 
with the expression known from quantum mechanics.

The total energy $\mathcal E=\mathcal E_{Pot}+\mathcal E_{Kin}$ 
also turns out to be identical for both charge distribution 
patterns (see Eqs. (\ref{2.1.10}) and (\ref{2.1.17})):
\begin{equation}
\label{2.1.18}
\mathcal E'=\mathcal E''=-\frac{1}{2}\cdot\frac{e^2}{a},
\end{equation}
which coincides with the results derived for a spherically 
symmetric $1S$ charge distribution (\ref{1.4.25}) and the 
appropriate formulas of quantum mechanics.

Calculate now the angular momenta for these two versions 
of charge distribution. We assume the charge to rotate about 
the $Z$ axis, with the velocity determined by Eq. (\ref{2.1.15}). 
The angular momentum is given by the formula
\begin{equation}
\label{2.1.19}
M_Z=\int mv_\varphi RdV,
\end{equation}
where $R=r\sin\vartheta=a\tau\sin\vartheta$ is the distance 
from the element of mass to the $Z$ axis. Formula (\ref{2.1.19})
 yields
\begin{equation}
\label{2.1.20}
M'_Z=\hbar\cdot0.998,\qquad M''=\hbar\cdot1.039.
\end{equation}
The integrals (\ref{2.1.19}) are expressed in terms of the 
gamma-function with a half-integer index, unlike the integrals 
of energy (\ref{2.1.16}) containing the gamma-function with an 
integer index.

Experiments suggest that the angular momentum of the $S$ state 
is $\hbar\cdot0.5$. This means that the values specified by Eq. 
(\ref{2.1.20}) disagree with the value known from experiment. 
One should therefore reconsider the process of charge/mass 
rotation in more detail.

\section{Potentials of a non-spherical charge distribution}

In Sec. 2.1, the energies of the $1NS$ state of the hydrogen 
atom are calculated under the assumption that a distributed 
charge does not have spherical symmetry. Experiment suggests, 
however, that the atom is spherically symmetric. It appears 
now reasonable to study what consequences would ensue from 
the assumption of the hydrogen atom being not spherically 
symmetric, and how such an atom would look to an observer. 
To do this, we have to calculate the potentials of distributed 
charges (\ref{2.1.1}) and (\ref{2.1.2}) and compare them with 
the Coulomb potential of the nucleus. The potentials of a 
distributed charge can be calculated from an expansion in 
spherical harmonics. The formulas 
pertaining to a search of a potential with the use of such 
an expansion can be found, for instance, in Ref. \cite{4}. 
As demonstrated in Sec. 2.1 of the present Paper, the 
distribution of the kind of Eq. (\ref{2.1.2}) can be expressed 
through two spherical harmonics, $Y_{00}$ and $Y_{20}$; therefore, 
the series will be truncated with only four terms left 
(two terms for $r>r'$ and two terms with $r<r'$). The 
function $\sin\vartheta$ being not a member of the $Y_{lm}$ 
system of spherical functions, the series for the potential 
is not truncated. As shown in Sec. 2.1, a charge distribution 
can be fully described by functions with $m=0$ only. Apart 
from this, these functions should be symmetric with respect 
to the equator. The
second condition suggests that a spherical function can have 
only even order $l$. The functions $Y_{l0}$ representing 
essentially Legendre polynomials were taken from Ref. \cite{6}. 
Five harmonics were taken for calculation of the potential: 
$Y_{00}$, $Y_{20}$, $Y_{40}$, $Y_{60}$, $Y_{80}$. In this case, 
the part of the multipole  moment which depends on 
the $Y_{80}$ function amounts to 0.01 of the moment depending 
on $Y_{00}$. The formulas employed in the calculation are 
specified in Appendix.

The main results of the calculations can be visualized in 
Figs. 2.1--2.6. The notation accepted is as follows. The radius 
of the spherical reference frame $\mathrm r$ is given in units 
of the Bohr radius $a$, i.e., $\mathrm r$ in the graphs is actually 
the parameter $\tau$ in all of the above formulas. The potentials 
$\mathrm U$ are expressed in units of $e/a$. The potential 
$\mathrm{U(r,}\vartheta)$ is the total potential deriving from 
the whole set of the harmonics involved. $\mathrm{U0(r)}$ is the 
potential deriving only from the spherically symmetric harmonic 
$Y_{00}$. Expressed in this notation, the Coulomb potential of 
a nucleus $\mathrm{N(r)}$ reads as $\mathrm{1/r}$.

\begin{figure}
\label{F1}
\begin{center}
\includegraphics[scale=0.2]{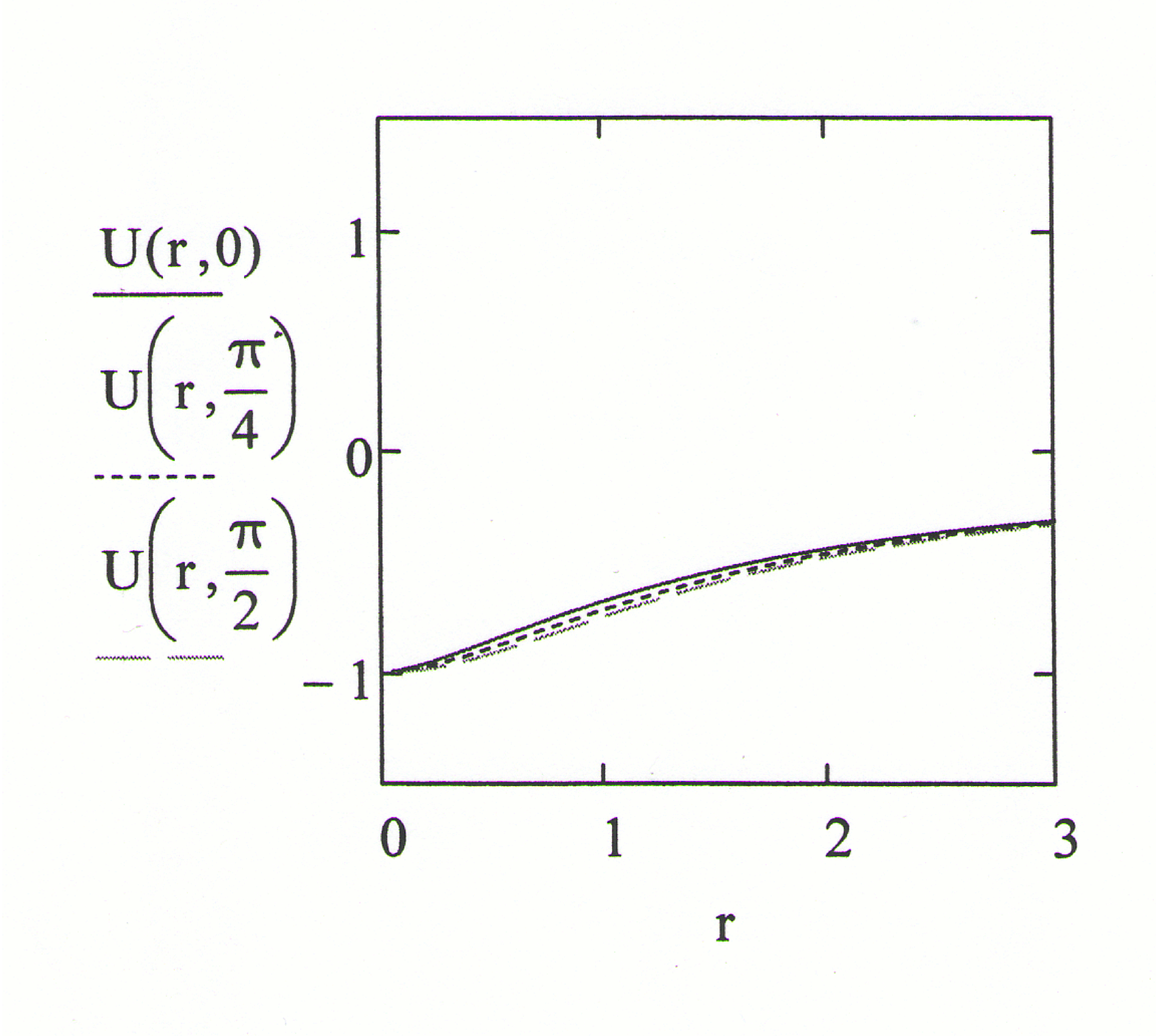}
\caption{}
\end{center}
\end{figure}

\begin{figure}
\label{F2}
\begin{center}
\includegraphics[scale=0.2]{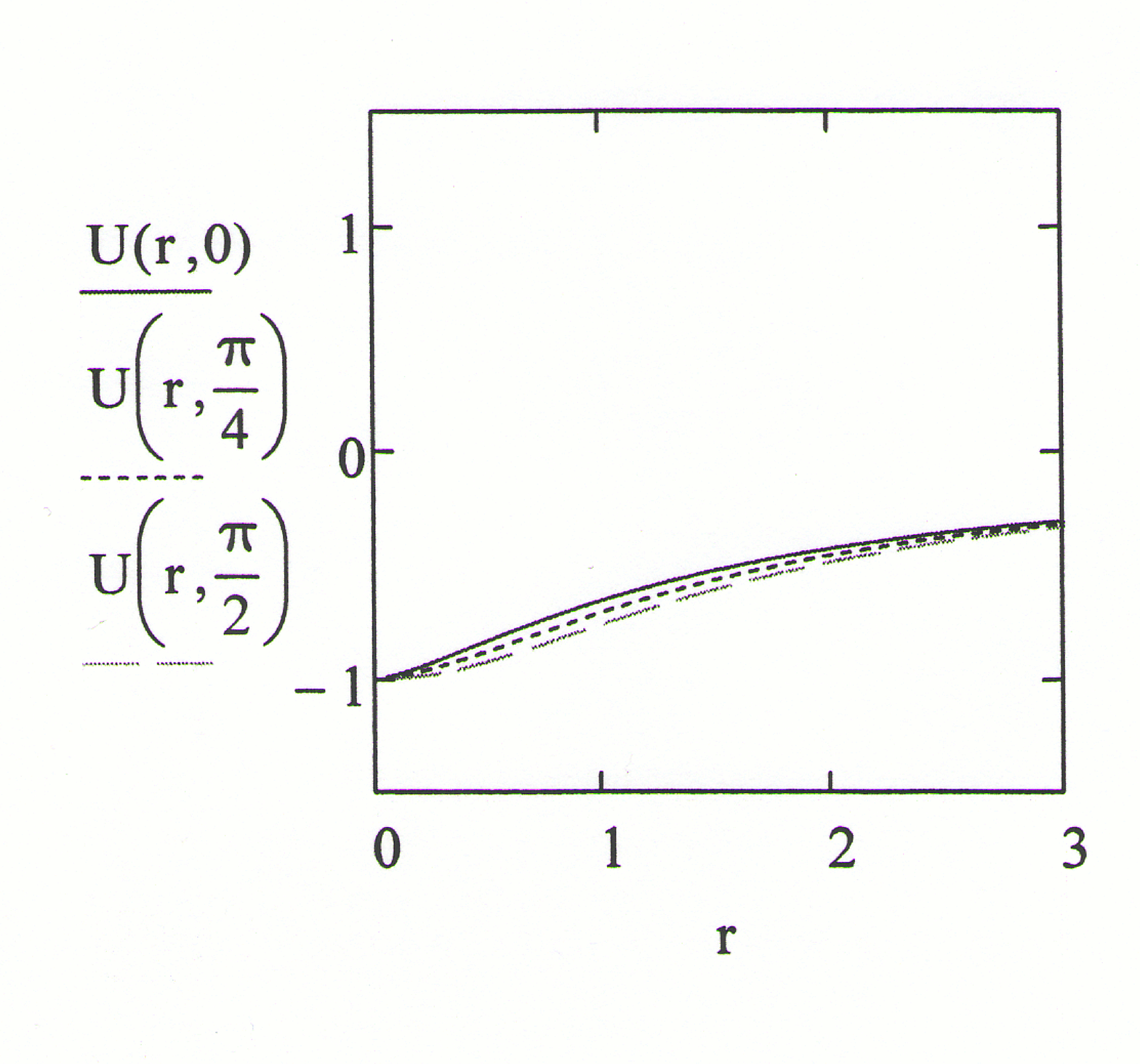}
\caption{}
\end{center}
\end{figure}

\begin{figure}
\label{F3}
\begin{center}
\includegraphics[scale=0.2]{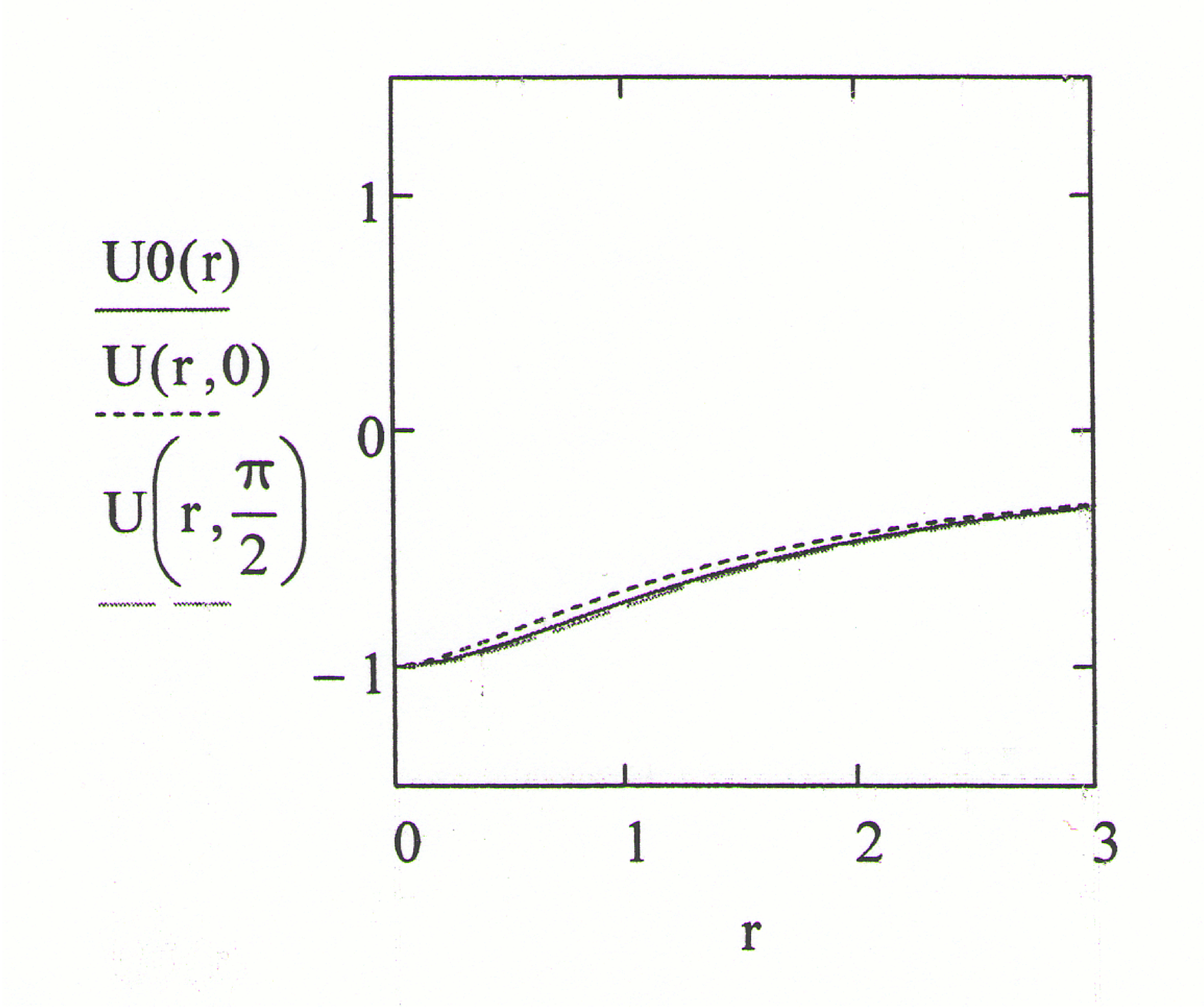}
\caption{}
\end{center}
\end{figure}

\begin{figure}
\label{F4}
\begin{center}
\includegraphics[scale=0.2]{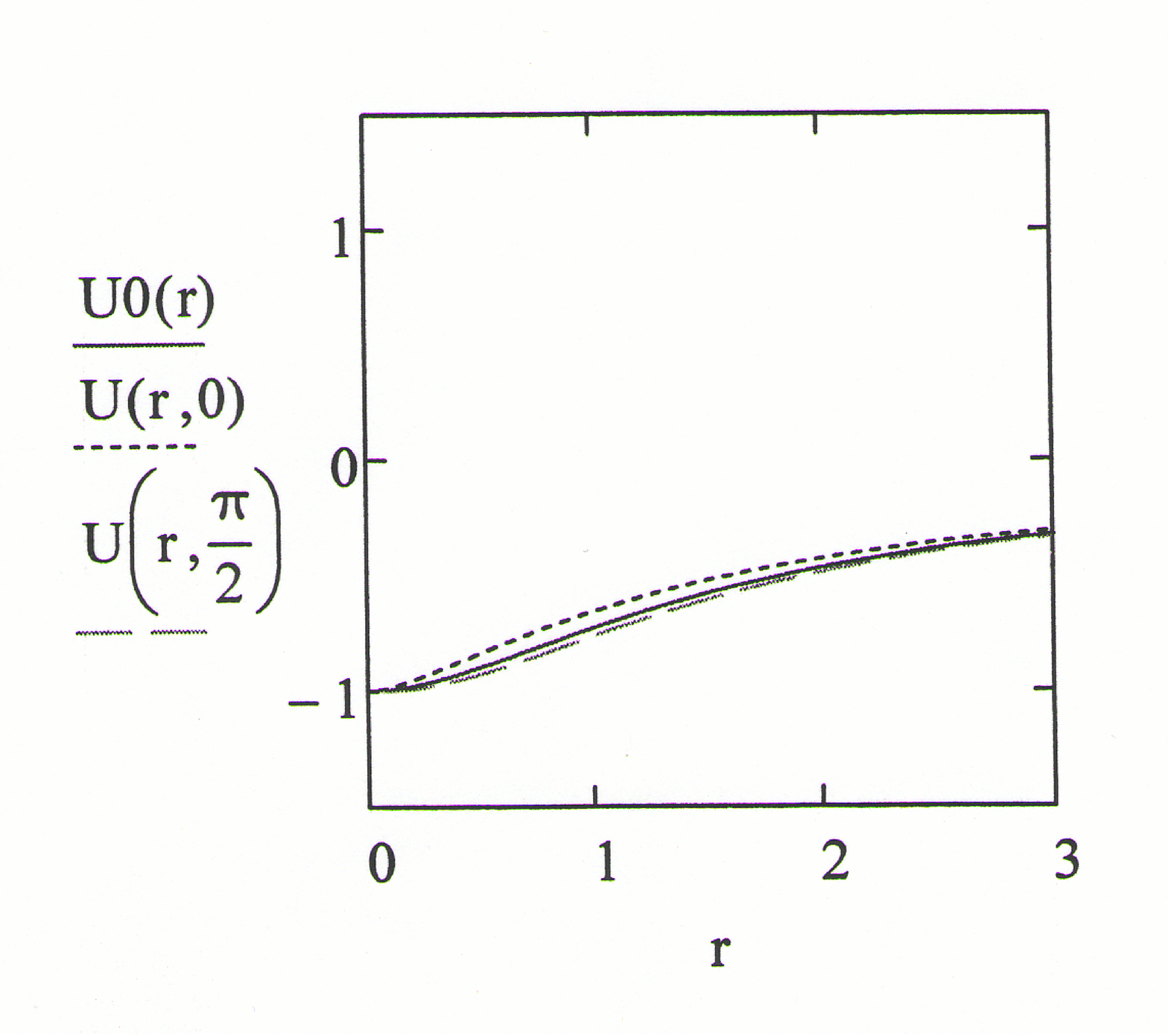}
\caption{}
\end{center}
\end{figure}

\begin{figure}
\label{F5}
\begin{center}
\includegraphics[scale=0.2]{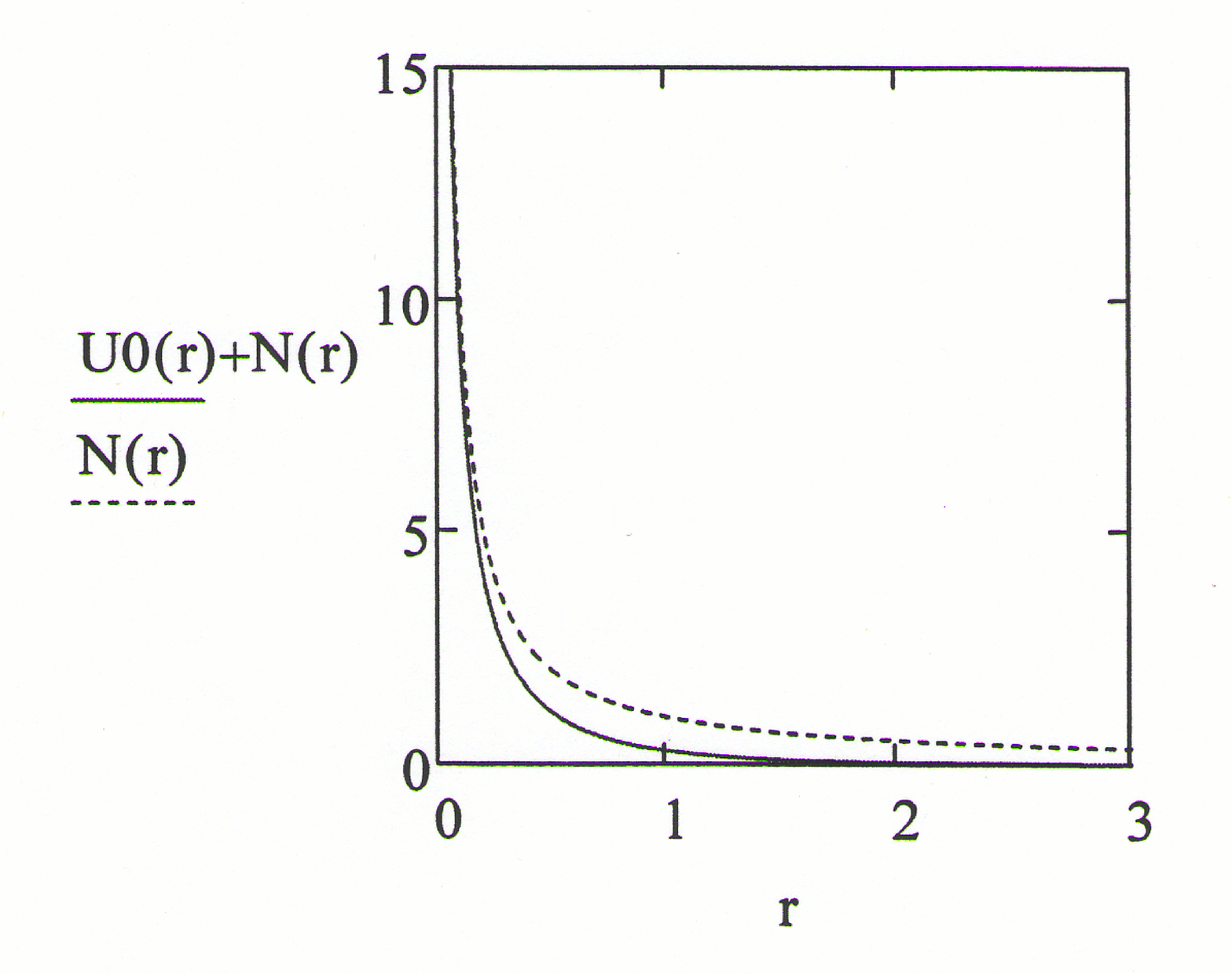}
\caption{}
\end{center}
\end{figure}

\begin{figure}
\label{F6}
\begin{center}
\includegraphics[scale=0.2]{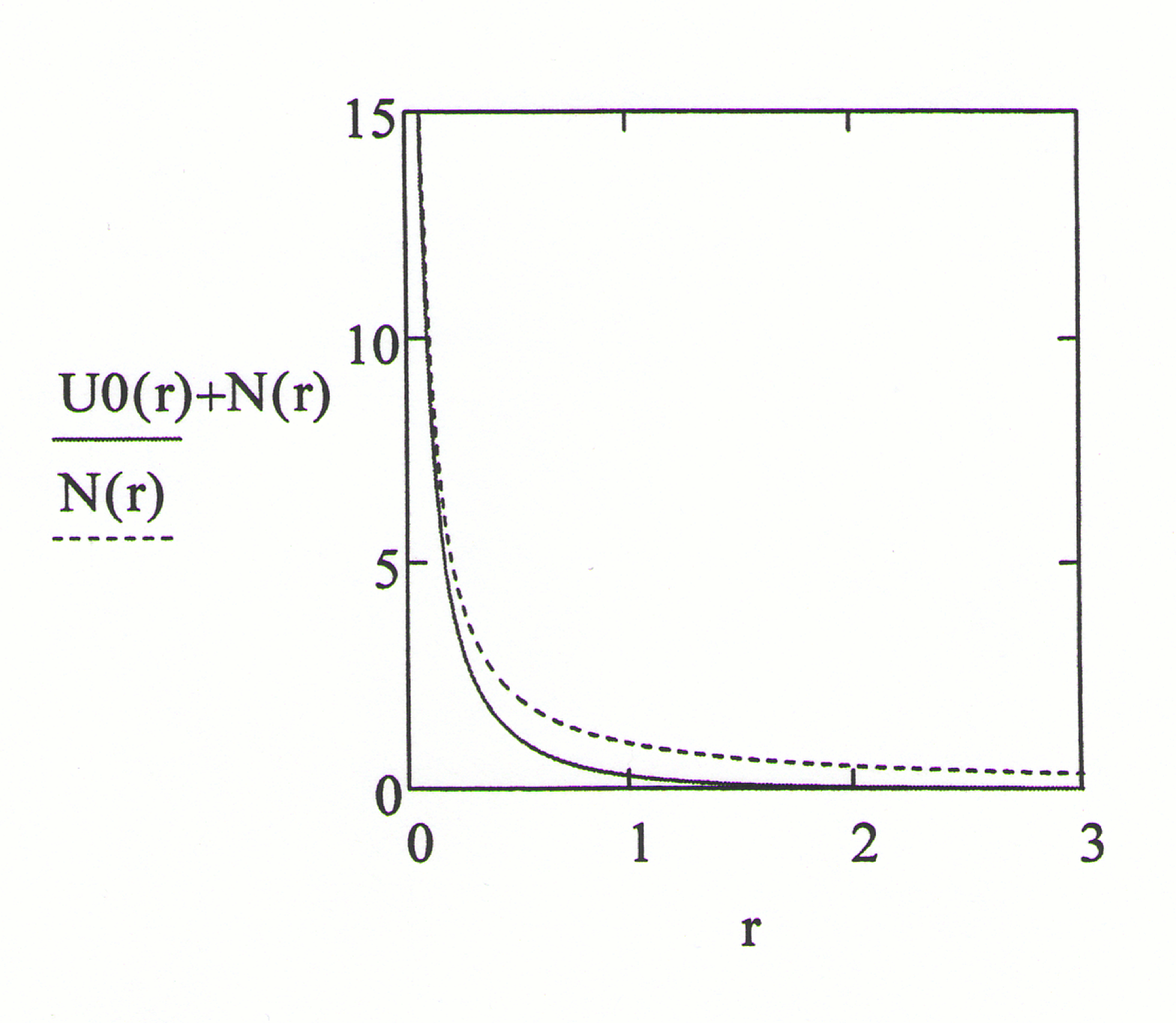}
\caption{}
\end{center}
\end{figure}

Figure 2.1 plots the potential of a distributed charge, 
Eq. (\ref{2.1.1}), vs. distance $\mathrm r$ from the nucleus 
for three angles: $\vartheta=0$, $\vartheta=\pi/4$, $\vartheta=\pi/2$. 
Shown in Fig. 2.2 is the same graph for the (\ref{2.1.2}) distribution. 
We readily see that the potentials depend very weakly on the angle 
$\vartheta$. And, as mentioned above, the potentials do not 
depend on the angle $\varphi$ at all. 

Figure 2.3 illustrates for the charge distribution (\ref{2.1.1}) 
the dependence on distance $\mathrm r$ from the nucleus of both 
the potential $\mathrm{U0(r)}$ formed by the spherically symmetric 
harmonic $Y_{00}$ only and of the total potential 
$\mathrm{U(r,}\vartheta)$ involving all the harmonics included 
for two angles, $\vartheta=0$ and $\vartheta=\pi/2$. Figure 2.4 
shows the same plot constructed for the charge distribution 
(\ref{2.1.2}). Examining these graphs we see that the potentials 
are formed in both cases mostly by the spherically symmetric 
harmonic, the contribution of the other harmonics being very small. 
This is immediately evident though from Figs. 2.1 and 2.2: indeed, 
if the potential is not small and the angular dependence is very 
weak, the only conclusion can be that the potential derives 
primarily from the spherically symmetric harmonic.

Thus, in spite of the charge distributions (\ref{2.1.1}) and 
(\ref{2.1.2}) being different from spherical, the potentials 
produced by these charges deviate very little from the spherical 
pattern. Therefore, {\itshape\bfseries such an atom would 
superficially look as spherically symmetric}. 

Consider now the extent to which the spherically symmetric 
harmonic of the potential of a distributed charge $\mathrm{U0(r)}$ 
affects the potential of the atom. An atom carries the total 
potential, i.e., the potential of the nucleus plus that of the 
distributed charge. Because the distributed charge potential 
primarily derives from the spherically symmetric harmonic 
$\mathrm{U0(r)}$, it is only this harmonic that we shall 
consider in this sum.

Figure 2.5 plots the total potential of a spherically 
symmetric harmonic of charge (\ref{2.1.1}) plus the Coulomb 
potential of the nucleus $\mathrm{N(r)}$. The dashed curve 
shows only the Coulomb potential of the nucleus $\mathrm{N(r)}$. 
We immediately see that for $\mathrm{r>2}$ the total potential 
is very small (which certainly derives from the fact that  
the field vanishes at some distance from the atom), but it 
approaches  the Coulomb potential 
of the nucleus as one comes closer to the latter. Figure 
2.6 plots the same graph for the charge distribution 
(\ref{2.1.2}). Significantly, the potential generated by 
the spherically symmetric harmonic $\mathrm{U0(r)}$ 
coincides with the potential $\Phi_{1S}$ of the spherical 
charge distribution (see Eqs. (\ref{1.3.7}) and (\ref{1.3.8})).

Thus, despite the absence of spherical symmetry in the charge 
distribution, the atom mostly preserves the main features of 
spherical symmetry; indeed, the deviation of the potential from 
the spherically symmetric pattern is very small, and as one 
comes closer to the nucleus, it approaches the Coulomb potential.

\section{Analysis of possible patterns of motion of a distributed 
charge}

As can be seen from Eqs. (\ref{2.1.20}), we have not obtained 
the correct value for the angular momentum of the hydrogen atom 
in ground state. This stresses the need for reconsidering the 
process of mass/charge rotation in more detail.

Equation (\ref{2.1.15}) shows essentially that each element of 
charge rotates circularly about the $Z$ axis. But why should 
each element of charge rotate in a circle only? Each element 
of the distributed charge is confined to the Coulomb potential 
well of the nucleus (we are disregarding as yet the additional 
potential of the distributed charge itself; as follows from 
Sec. 2.2, the difference of the total from the Coulomb potential 
in an atom is small). The behavior of a charge in a Coulomb 
potential well is known. Monograph \cite{7} could be best 
suited for our purposes.

Consider the situation in two stages. We shall first be 
interested in the motion of a charge at the equator.

Recall some well established facts.

In a Coulomb potential well, a constant element of charge 
$dq$ with a mass $dm$ can move along a circle or an ellipse, 
and the ellipse can degenerate into a straight line. The 
energy of an element of charge depends on the semimajor axis 
of the ellipse (or on the radius of the circle). The actual 
shape of the ellipse (i.e., its semiminor axis) depends on 
the angular momentum of the particle. Thus, all ellipses with 
the same semimajor axis but different semiminor axes have the 
same energies but different angular momenta, down to the zero 
momentum (in which case the ellipse degenerates into a straight 
line). Said otherwise, mass/charge elements of the same energy 
can move in a Coulomb potential well along trajectories which 
differ in the value of the angular momentum.

Because we have a distributed electron charge in an atom, it 
appears only logical to assume that each element of charge 
can move along any allowed trajectory (by an allowed trajectory 
we understand here any trajectory satisfying the laws of mechanics). 
Note, however, that different trajectories (different ellipses) 
intersect. Therefore, when introducing the assumption that 
elements of charge can move along different allowed trajectories, 
we have to accept another one as well, namely, that each element 
of mass/charge can move along its trajectory regardless of those 
of other charges. In other words, charges {\itshape\bfseries may 
pass through} one another without an attendant change of the 
trajectory. Said otherwise, each element of mass/charge moves 
in the force field independently of other elements of mass/charge. 
It goes without saying, that all elements surrounding the element 
under consideration contribute to the force field.

This would seem to be contrary to the observation that 
charges interact with one another. But the statement that 
each mass/charge element moves in a force field independently 
of other mass/charge elements is just a consequence of the 
fact that we consider interaction of charges not directly 
with one another but rather through the field; indeed, each 
charge interacts with the field generated by another charge 
or by all the other charges. In other words, we consider 
charge motion in a force field created by other charges, 
putting the existence of these charges apart.

Like charges repel one another. Therefore, the assumption 
that likely charged elements of charge can pass through 
one another may seem a far-fetched idea. Indeed, the point 
charges one usually considers have an infinite density at 
the point where the charge is located. Therefore, like 
point charges cannot pass through one another; more than 
that, they cannot even approach one another close enough. 
The distributed charges treated by us here have a finite 
charge density. Such charges can penetrate into one another, 
depending on what external forces act on these charges and 
what are the forces created by these charges. 

In actual fact, the statement that charges can pass through 
one another does not carry anything supernatural in it. For 
instance, electromagnetic fields can penetrate one into or 
through the other without at the same time affecting one 
another---this is nothing but the standard principle of 
superposition. Two radar beams can cross without interaction; 
this is just penetration of ac fields through one another. 
Superposition of one dc field on another (the principle of  
superposition) may be regarded as penetration of one field 
into another. Significantly, in this process the fields do 
not act in any way on one another.

As for the charges, no statements concerning passage of one 
charge through another without direct action on one another 
(interaction of charges is taken into account through the 
fields created by these charges) have thus far been made, 
although the principle of superposition is valid for charges 
as well. This statement should, however, be made. If 
electromagnetic fields do pass through one another, there 
would appear nothing strange in admitting that charges 
likewise can do it. The difference between these statements 
lies in that ac electromagnetic fields propagate along 
rectilinear trajectories (trajectories (beams) are straight 
lines (in vacuum)), while charges move along their 
trajectories in a potential field. The actual shape of the 
trajectory is determined by the potential field in which 
this element moves, as well as by the parameters of this 
element. A trajectory can be calculated in the frame of 
theoretical mechanics. In our case, the trajectories along 
which an element of distributed charge/mass moves in the 
field of the nucleus are closed curves rather than straight 
lines.

Consider in more detail the motion of an element of charge 
in an atom.

Assume an element of charge $dq$ located in the equatorial 
plane. Consider the trajectories in moving along which the 
charge $dq$ has the same total energies (in this case all 
elliptical trajectories have the same semimajor axes). This 
element can move in a circular or an elliptical trajectory 
in the equatorial plane, with the total energy of this 
element in any trajectory being the same, and only angular 
momenta different (see, e.g., Ref. \cite{7}). Each angular 
momentum can be identified with its own elliptical trajectory. 
Because in all trajectories the element of charge $dq$ has 
the same energy, this element of charge can move 
{\bfseries\underline {along any}} trajectory. Moreover, 
this element of charge can move {\bfseries\underline {in all 
trajectories at the same}} {\bfseries\underline {time}}. This 
can be visualized in the following way. Divide element   
of charge $dq$ in $k$ parts. Then one element of charge 
$dq'=dq/k$ can move along one elliptical trajectory, another 
charge $dq'$, along another trajectory, and so on. As $k$ 
tends to infinity, all the trajectories will criss-cross 
all of the allowed region containing trajectories of the 
elements of charge $dq'$ of the same energy but with 
different angular momenta. Generally speaking, this 
process may be considered not as motion of elements of 
charge along trajectories but rather as motion of a 
continuous medium, of a {\itshape\bfseries charge wave}. 

Let us analyze the various trajectories along which an 
element of charge $dq$ with a mass $dm$ can move in the 
case where the total energy of the element in each trajectory 
is the same. 

\begin{figure}
\label{F7}
\begin{center}
\includegraphics[scale=0.2]{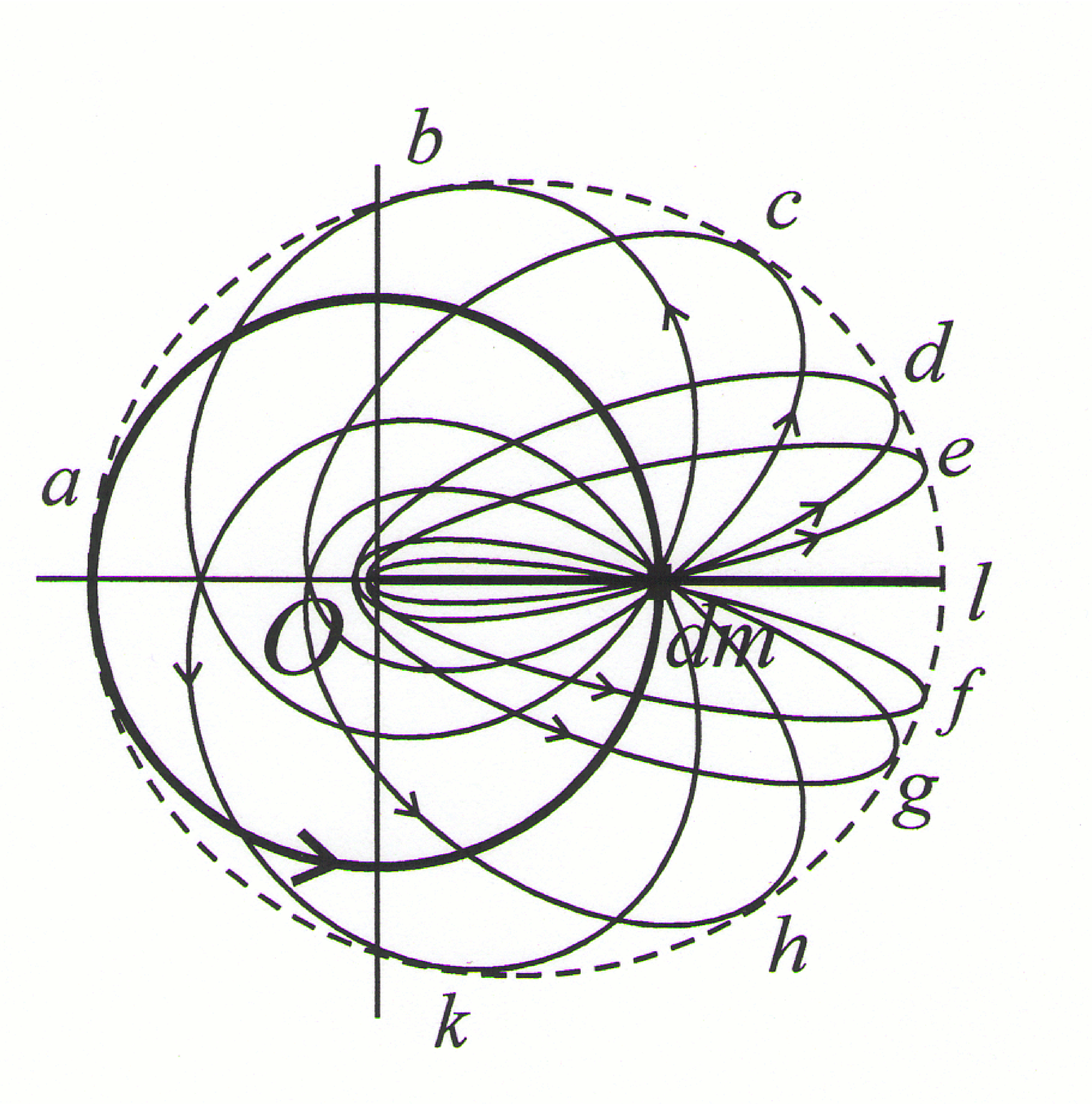}
\caption{}
\end{center}
\end{figure}

Figure 2.7 illustrates several such orbits of all possible 
ones: a circular orbit $a$,  eight elliptical orbits 
$(b-k)$ with different eccentricities (and, hence, different 
angular momenta), and a linear orbit $l$ into which the 
ellipse degenerates at an eccentricity of unity. This orbit 
passes through the nucleus of the atom. All the orbits are 
characterized by identical semimajor axes (if the energies 
of the elements are equal, the semimajor axes of the ellipses 
should likewise be equal). All orbits lie in the same plane. 
The elements of charge in all orbits rotate in the same 
direction.

All orbits focus at the same point. In this focus (in our 
figure, this is the center $O$ of the circle) the nucleus 
of the atom is located. Using the focal properties of 
ellipses, one can readily show that each elliptic trajectory 
intersects a circular orbit at the point where this ellipse 
intersects its semiminor axis. The dashed lines confine the 
region of allowed trajectories along which an element $dm$
can move.

As already mentioned, an element of charge $dq$ with a mass 
$dm$ can move along all of the above trajectories simultaneously 
to form not propagation of single particles but rather a wave 
motion. For this to become possible, the element of charge 
$dq$ has to split into a multitude of parts $dq'$, each of 
them moving in its own trajectory.

An element of mass/charge residing in a Coulomb potential well 
moves in one plane. We have considered motion in the equatorial 
plane. Note, however, that through a line connecting the element 
under study with the nucleus one can pass an infinite number of 
planes. The trajectory of a given element may lie in any of these 
planes, because Coulomb field possesses spherical symmetry. 
Moreover, this element can be divided into a multitude of parts, 
and each part of the element can move in a trajectory in its 
plane. It thus appears that the element of charge under 
consideration can move along all of the allowed trajectories in 
all planes simultaneously. It would apparently be more 
appropriate to speak here not of the motion of a set of 
mass/charge elements but rather of that of a wave propagating 
within a certain {\itshape\bfseries solid  angle}. The solid 
angle is defined by the set of all allowed trajectories. 

This reasoning can be repeated for any element of charge. 
Each element can be divided into parts, and these parts 
will propagate in a certain solid angle. What we will have 
actually is propagation within a certain solid angle of a 
{\itshape\bfseries charge wave}, or, to express it more 
properly, a {\itshape\bfseries mass/charge wave}.

This reasoning resembles in a large measure the 
{\itshape\bfseries Huygens--Fresnel principle}. By the 
Huygens--Fresnel principle, each point of a propagating 
wave acts as a source of a secondary wave, and the front 
of the propagating wave may be visualized as an envelope 
of all the secondary waves.

The difference of the consideration offered here from the 
Huygens--Fresnel principle lies essentially in that by the 
latter principle a wave propagates along a straight line, 
i.e., the rays of all secondary waves are straight lines. 
In our case, the trajectories associated with the propagation 
of charge waves, rather than being straight lines, are 
mediated instead by the potential of the field in which the 
element under consideration moves and by the parameters of 
the element itself. In the cases of interest here, these 
trajectories are closed curves.

The similarity with the Huygens--Fresnel principle lies in 
that any point of the distributed charge is a center from 
which a mass/charge wave propagates within a certain solid 
angle.

The above pattern may be considered as an attempt to find 
common features between the corpuscular and wave patterns 
of behavior. There are even grounds to suggest that the 
corpuscular and wave concepts actually merge. The grounds 
underlying this statement may be seen in that the behavior 
of a distributed charge can be studied in two different 
contexts. The behavior of each single mass/charge element 
obeys all laws of theoretical mechanics. Its motion can be
 calculated with the use of equations derived in theoretical 
 mechanics. On the other hand, when one considers the motion 
 of all elements taken together, it is the motion of a wave.
 This wave should obey certain partial differential equations  
 involving the effect of potentials on the motion of the 
 “charge waves”. These equations have not thus far been 
 constructed. But it is with partial differential equations that  
the motion of “charge waves” is most appropriate to analyze, 
and this stresses the need for constructing such equations. 
When such equations allowing for the effect of potentials on 
“charge wave” motion are obtained, there will be strong grounds 
to call the field of science described by these equations 
{\itshape\bfseries wave mechanics}, because these equation 
should take into proper account both the wave properties of 
objects (originating from specific boundary or periodic 
conditions) and all the characteristics described by 
theoretical mechanics. Until this is done, the term 
“wave mechanics” announced in the title of the Paper 
should be treated rather as an expression of wishful 
thinking on the part of the author.

One should also attempt to apply the enormously vast 
amounts of knowledge amassed in theoretical mechanics 
for point objects to description of the behavior of a 
continuous medium. We have to admit, however, that the
 behavior of a continuous medium could be described more 
 adequately by the “wave” formalism, i.e., through the 
 use of partial differential equations.

 \section{Angular momentum of the $1NS$ state}
 
We turn now to calculating the angular momentum of the 
ground state of a hydrogen atom. We start by dividing the 
volume of the distributed mass into elements of magnitude 
$dV$ with a mass $dm=mdV$. Next we calculate the angular 
momenta of each element separately and add them subsequently. 
We have to keep in mind that each element of mass $dm$ can 
move along different elliptical orbits. Therefore, we have 
to consider in the beginning  the angular momentum of one 
element of mass.
 
A Coulomb potential well is spherically symmetric. The 
presence of a distributed charge, which is anything but 
spherically symmetric, distorts the symmetry of this 
potential. As shown in Sec. 2.2, however, the presence of 
this charge affects very little the potential. This gives 
us grounds to assume in what follows that elements of 
charge move in a Coulomb potential field. 

Calculate the angular momentum of a charge/mass element 
subject to the condition that the element moves along 
different orbits in the same plane but that in all these 
orbits the energy of the element is the same. We again 
use the data given in Ref. \cite{7}. The angular momentum 
$dM$ of a constant element of mass $dm$ in a Coulomb 
potential well can be written as
\begin{equation}
\label{2.3.1}
dM=\frac{2dmf}{T},
\end{equation}
where $f$ is the area of the orbit, and $T$ is the period 
of revolution of an element of mass in this orbit. Recall 
that the period $T$ depends only on the energy of this 
element with the mass $dm$. Because we consider here orbits 
of the same energy, the value of $T$ for all the orbits of 
interest will be the same. In a Coulomb potential well, 
orbits are actually ellipses; therefore, we obtain 
$f=\pi xy$, where $x$ is the semimajor, and $y$, the semiminor 
axes of the ellipse. Because elements of mass in all the 
orbits of interest to us here have the same energy, the 
semimajor axes $x$ of all the orbits are identical (the 
semimajor axis depends on energy only), and the semiminor 
ones, $y$, are different.

An element placed in a Coulomb potential well moves in one 
plane only. Consider the motion of an element in one of
 such planes.

If an element $dm$ rotates in a {\itshape\bfseries circular} 
orbit of radius $R=x$, for the angular momentum of this 
element we can write
\begin{equation}
\label{2.3.2}
dM_R=\frac{2dm\pi x^2}{T}=\frac{2dm\pi R^2}{T}.
\end{equation}
Calculate now the angular momentum for the case where an 
element of mass $dm$ moves in this plane along {\itshape\bfseries 
all orbits at the same time}. To do this, divide the element 
of mass $dm$ into $k$ parts: $dm'=dm/k$ (see Fig. 2.7).

Each element of mass $dm'$ will move along its elliptical 
trajectory. For its angular momentum we can write (recall 
that $x=const=R$ for this energy):    
\begin{equation}
\label{2.3.3}
dM'=\frac{2dm'\pi xy}{T}=\frac{2dm'\pi Ry}{T}.
\end{equation}
To calculate the total angular momentum $dM_{dm}$ of an 
element $dm$ moving in all trajectories in the plane under 
consideration simultaneously, we have to sum all the momenta 
$dM'$ (see Eq. (\ref{2.3.3})). Significantly, the parameter 
$y$ varies in the process from zero (the case in which the 
ellipse degenerates into a straight line) to $R$ (where the 
ellipse transforms into a circle of radius $R$). The element 
of mass $dm'$ can be prudently recast to the form 
$dm'=dm\frac{dy}{R}$, because the quantity $R/dy$ is nothing 
else but the number of parts $k$ into which the mass $dm$ was 
divided. But then the total angular momentum becomes
\begin{equation}
\label{2.3.4}
dM_{dm}=\frac{1}{T}\int\limits_0^R 2\pi ydmR\frac{dy}{R}
=\frac{1}{T}2\pi dm\int\limits_0^R ydy=\frac{1}{T}2dm\pi 
R^2\cdot\frac{1}{2}.
\end{equation}
Examining Eqs. (\ref{2.3.2}) and (\ref{2.3.4}), we see 
that the angular momentum of an element of mass $dm$, in 
the case where it moves in all trajectories simultaneously, 
is only one half that of the element of mass $dm$ moving 
in a circular orbit as a whole (the energies of the $dm$ 
elements are in both cases the same). In other words, 
calculation of the angular momentum of an element $dm$ 
moving along all allowed elliptical orbits may be replaced 
by calculation of the angular momentum of the same element
 but moving along a circular orbit. The necessary condition 
 for this to be valid is that the total energies of the 
 element in the circular and elliptical orbits should be 
 equal (indeed, in both orbits we have the same element). 
 For this to be valid, the semimajor axes of all the ellipses 
 considered should be equal to one another and to the radius 
 of the circle. It may be appropriate to recall that we are 
 speaking here about trajectories confined to one plane. 
 Thus, Eq. (\ref{2.3.4}) can be recast in the form
\begin{equation}
\label{2.3.5}
dM_{dm}=dM_R\cdot\frac{1}{2}.
\end{equation}
Here $dM_R$ is the angular momentum of the element $dm$ rotating 
along the {\itshape\bfseries circle} of radius $R$ in the given 
plane. $dM_{dm}$ is the angular momentum of the element $dm$ 
moving in {\itshape\bfseries all allowed trajectories}, likewise 
in the same plane, with the elements in both cases having equal 
energies.

We calculated earlier the angular momentum of the distribution 
of charges rotating about the $Z$ axis (Eqs. (\ref{2.1.19}) and 
(\ref{2.1.20})). In this case, each element of charge was rotating 
in circular orbits whose planes were parallel to the equatorial 
plane. One just could not conceive at the time of any other 
pattern of rotation for a charge distribution. In this version 
of rotation, it was difficult to identify, however, the mechanism 
accounting for rotation of charges lying outside the equatorial 
plane, because the plane of their orbits does not pass through 
the center of force, i.e., the nucleus.

Having allowed for the possibility of charges interpenetrating 
one another, we could construct a different pattern of rotation, 
which would appear more natural while not contradicting any laws 
of mechanics. An element of charge, acted upon by the force of 
attraction, moves in the Coulomb potential well of the nucleus.
 This element rotates in actual fact about the nucleus rather 
 than about the $Z$ axis. The trajectories of motion lie in the 
 plane crossing the point where the nucleus is located, i.e., 
 the origin of the coordinate frame. In this case, the trajectories, 
 rather than being parallel to the equatorial plane, can make any 
 angle with it.

A Coulomb potential well is spherically symmetric. As a consequence,
 the orbit of an element $dm$ may lie in different planes. 
 These planes can be visualized by rotating the original 
 plane about the line connecting the position of the element 
 $dm$ with that of the nucleus. The orbit of an element $dm$ may 
 lie in any of these planes. Moreover, because the energies of 
 the element in each trajectory are equal, the element $dm$ may 
 move along elliptical trajectories in all these planes at the 
 same time. It appears only natural that, as already mentioned, 
 in actual fact one should treat this pattern as motion of a 
 charge wave within a certain solid angle rather than as that 
 of elements.

\begin{figure}
\label{F8}
\begin{center}
\includegraphics[scale=0.2]{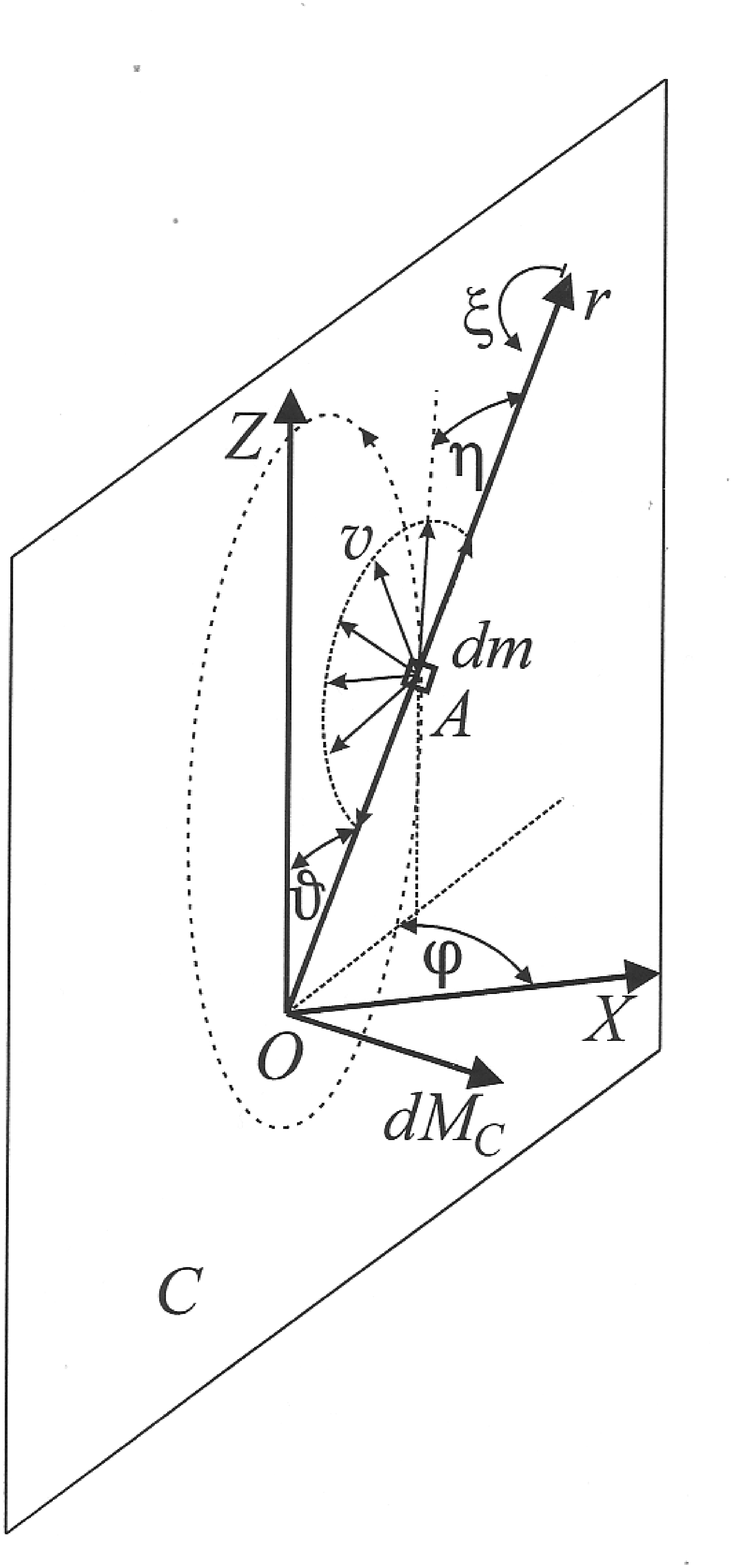}
\caption{}
\end{center}
\end{figure}

Consider this situation in more detail. We take first the 
trajectories lying in the plane passing through the vectors 
$Z$ and $r$ (vector $r$ specifies the direction to the element 
chosen). We will call it plane $C$ (Fig. 2.8). Let the element 
be located at point $A$. Figure 2.8 displays a “fan”\ of 
velocities $v$, i.e., directions along which an element $dm$ 
can move in the $C$ plane (compare with Fig. 2.7). Recall that 
all elements rotate in one sense. The direction of the 
velocities of element $dm$ at point $A$ is defined by that 
of the tangents to the elliptical trajectories at point $A$. 
One of such trajectories, i.e., one of the ellipses is 
identified in Fig. 2.8 with a dashed line. The direction 
of motion of an element can be described by the angle $\eta$, 
which we will reckon from the $r$ axis. The angle $\eta$ 
defines the direction of motion of an element along the 
trajectory, i.e., along the ellipse. This angle varies from 
$0$ to $\pi$, which corresponds to variation of the ellipse 
shape from a straight line to a circle and again back to the 
straight line (see Fig. 2.7).

Because the velocity fan lies in the $C$ plane, the resultant 
velocity formed by summation of the velocities of the element 
$dm$, which propagates along different trajectories in the $C$ 
plane, at point $A$ lies in the same $C$ plane, while the 
angular momentum $dM_C$ of the element $dm$ is perpendicular 
to this plane (see Fig. 2.8).

\begin{figure}
\label{F9}
\begin{center}
\includegraphics[scale=0.2]{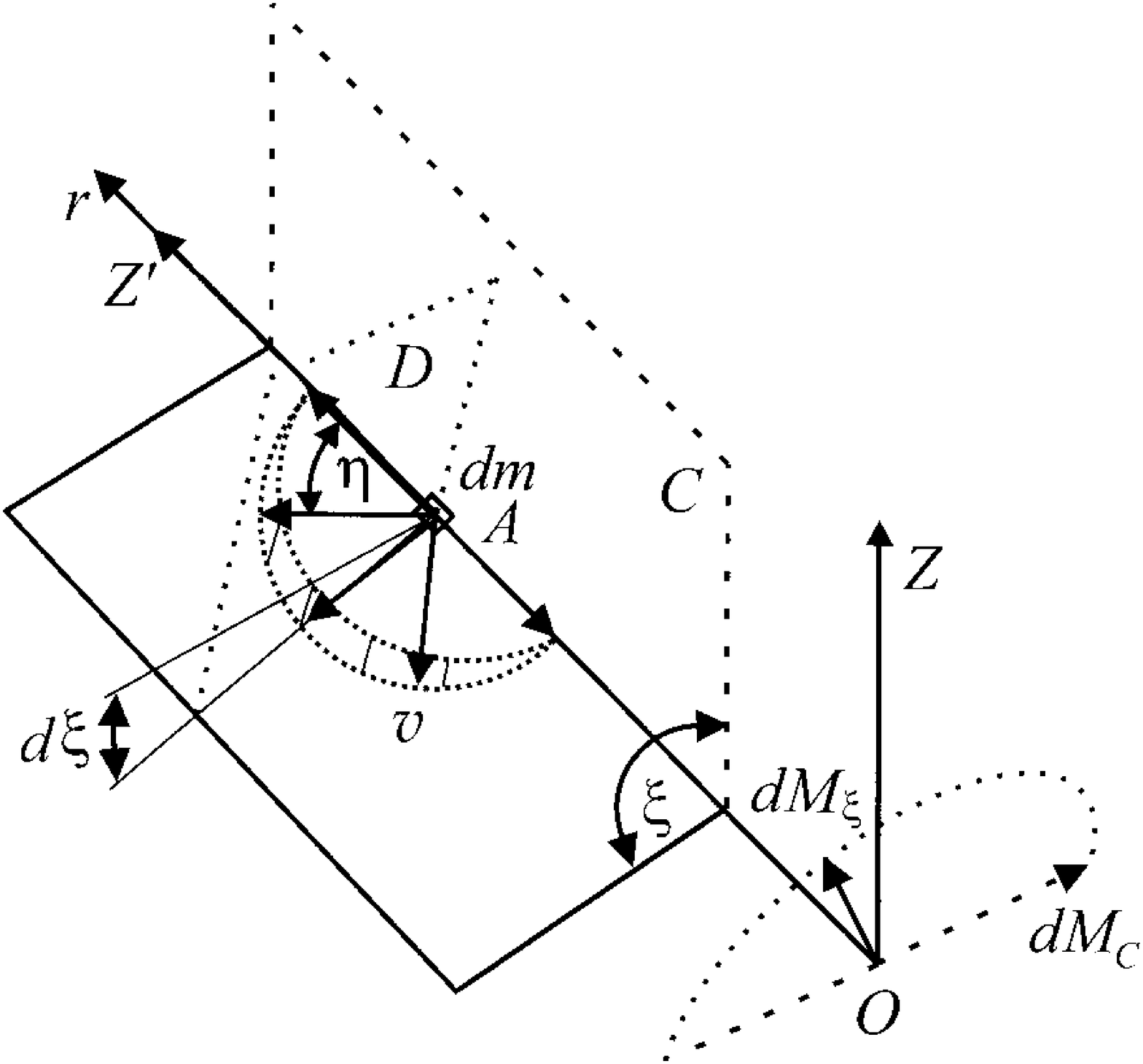}
\caption{}
\end{center}
\end{figure}

The trajectories of the element $dm$ may lie not only in plane 
$C$. The other planes to which the trajectories may be confined 
can be obtained by rotating plane $C$ about the axis connecting 
the origin $O$ with the position of the element $dm$, i.e., 
point $A$ (Fig. 2.9). We denote the angle of turn of this plane 
by $\xi$, and will reckon it from the original position of the 
$C$ plane. The reason for which this angle is reckoned from this 
plane will become clear later.

The angular momentum of an element $dm$ moving in any turned plane 
is perpendicular to this plane. Therefore, as the planes are turned, 
the tips of the momentum vectors (we will denote them by $dM_\xi$) 
will trace an arc. This arc will lie in a plane perpendicular to 
vector $r$ (Fig. 2.9).

Consider the limits within which the angle $\xi$ can vary. Variation 
of the angle $\xi$ will initiate formation of new planes over which 
the element $dm$ can move. We have to keep in mind that the element 
can move not only in these, newly formed, planes, but in all planes 
simultaneously. We readily see that if the angle $\xi$ is larger 
than $\pi$, new planes coinciding with some of the original planes 
will appear. In these coinciding planes, all allowed trajectories 
will coincide as well. Significantly, motion over these coinciding 
trajectories will occur simultaneously in opposite senses, i.e., 
there will be no motion on these planes. (This can be readily seen 
from Fig. 2.8 if we turn mentally plane $C$ through $\pi$, or 
from Fig. 2.9.)

Thus, angle $\xi$ can vary from $0-\pi$. 

As a plane turns through $\pi$, vectors $dM_\xi$ likewise turn 
through $\pi$.

Consider now why the angle $\xi$ should be reckoned from the $C$ 
plane.

The overall rotation of the charge in an atom (by overall 
rotation we understand here rotation of the charge as a whole 
rather than that of individual elements of charge) occurs in 
one sense, which accounts for the atom having an angular 
momentum. We conventionally directed the angular momentum of 
the atom along the $Z$ axis. Accordingly, the resultant 
velocity of rotation is directed along the $\varphi$ axis.

Now if the trajectories of an element lie in the $C$ plane, 
all components of the velocity lie in the same plane, with 
no velocity components left along the $\varphi$ axis. The 
angular momentum $dM_C$ of this element will in this case 
have no components along the $Z$ axis (see Fig. 2.8). If we 
rotate the $C$ plane by increasing the angle $\xi$, we will 
detect formation in these turned planes velocity components 
directed along the $\varphi$ axis, with the $dM_\xi$  momentum 
acquiring a component along the $Z$ axis (see Fig. 2.9). 
Both quantities reach a maximum for $\xi=\pi/2$. As the angle 
$\xi$ increases still further, both quantities will decrease, 
to vanish eventually at $\xi=\pi$.

As the angle $\xi$ grows still more (as does the corresponding 
turn of the plane), the elements will rotate in these planes 
in the opposite sense, with a negative component of the velocity 
along the $\varphi$ axis, and of the angular momentum along the 
$Z$ axis, appearing, although they should not exist by our 
original condition.

Thus, reckoning the angle $\xi$ from the $C$ plane and variation 
of $\xi$ within the $0-\pi$ limits provides overall rotation 
of the charge in one sense and formation of an angular momentum 
along the $Z$ axis. Significantly, no negative components of 
the angular momentum appear along the $Z$ axis.

The motion of elements over certain planes we have just considered 
is only some approximation to reality, because strictly confined 
to one plane is only motion of an element of a constant magnitude. 
Now in a real atom an element of mass/charge propagates within a 
certain solid angle. We will have to calculate now the angular 
momentum of an element $dm$ in the case of its propagation within 
a solid angle. We start with constructing a local frame of 
spherical coordinates $r'$, $\eta$, $\xi$ centered on the element 
$dm$, i.e., at point $A$. The $\eta=0$ axis will be directed 
along the $r$ axis and will be called the $Z'$ axis, and the 
angle $\xi$ will be reckoned from plane $C$ (Fig. 2.9). We see 
immediately that the angles $\eta$ and $\xi$ of this coordinate 
system coincide with the angles $\eta$ and $\xi$ considered above.

If an element propagates into a solid angle, this means actually 
that its trajectory, rather than being confined to a certain 
plane, occupies a sector instead. An analog of element motion 
in a plane will be motion within a small solid angle $d\xi$, 
the orientation of this solid angle $d\xi$ (an analog of the 
position of the plane) being determined by the angle $\xi$. 
As already demonstrated, the angle $\eta$ varies within the 
$0-\pi$ limits, and the angle $\xi$ varies within the same 
limits, $0-\pi$. Thus, the element propagates into a 
hemisphere; accordingly, the solid angle into which the 
charge propagates as a wave is $2\pi$.

Our problem lies in finding the resultant angular momentum 
of the element $dm$ which propagates into a solid angle as 
a wave, for which purpose one will have to sum all components 
of the angular momentum oriented in different directions. 
Significantly, any direction of velocity at a given point (and 
within the allowed solid angle) is equally probable. This 
conclusion is valid because an element at a given point which 
propagates in different directions has the same energies, i.e., 
all these trajectories are equally probable. Moreover, the 
magnitude of the velocity should not be dependent on the 
direction of motion of a given element. This conclusion can 
be substantiated in the following way. We consider ellipses 
of the same energy, i.e., the total energy of an element on 
any ellipse is the same (although the relative magnitudes of 
the kinetic and potential energies change as the element 
moves along the ellipse). The potential energy depends on 
the position of the element only. Hence, at a given point 
(for instance, at point $A$) the potential energies of an 
element moving over any ellipse are the same. But if the 
total energies are equal, and the potential energies at a 
given point are equal too, then the kinetic energies at 
this point will be equal as well. For this reason, the 
velocities are equal irrespective of their direction. To 
calculate the angular momentum of the element $dm$ propagating 
into a solid angle $2\pi$, we divide the element $dm$ into parts
\begin{equation}
\label{2.3.6}
dm'=dm\frac{d\Omega}{2\pi}.
\end{equation}

Each element $dm'$  propagates into a solid angle 
$d\Omega=\sin\eta d\eta d\xi$.

We shall approach this problem in steps. Isolate a sector $d\xi$. 
Find the projection of the angular momenta of the elements 
propagating into the $d\xi$ sector onto a plane perpendicular 
to the $r$ axis (plane $D$ in Fig. 2.9). To do this, we will 
have to sum the angular momenta over the coordinate $\eta$. 
Because at a given point the velocities in any direction are 
equal in magnitude, the distribution of the angular momenta 
in the $d\xi$ sector is symmetric relative to the $D$ plane. 
Denoting this projection by $dp_\xi$, we come to 
\begin{equation}
\label{2.3.7}
dp_\xi=\int v\sin\eta dm'=\int v\sin\eta dm\frac{d\Omega}{2\pi}=
\int_0^\pi dm\frac{\sin\eta d\xi}{2\pi} v\sin\eta d\eta=
\frac{dmv}{2\pi}\cdot\frac{\pi}{2}\cdot d\xi.
\end{equation} 

The angular momentum $dM_\xi$ corresponding to the momentum 
$dp_\xi$ is shown in Fig. 2.9.

We turn now to summation of the vectors $dp_\xi$ obtained over 
the coordinate $\xi$. We first find the projection of vectors 
$dp_\xi$ on the plane formed by turning plane $C$ through an 
angle $\xi=\pi/2$. Because at a given point the velocities in 
any direction are equal, the distribution of the momenta 
$dp_\xi$ will be symmetric relative to this plane. The momenta 
$dp_\xi$ lying in plane $D$ perpendicular to vector $r$, the 
vector we have obtained (denote it by $dp_\varphi$) will 
coincide in direction with the coordinate $\varphi$:
\begin{equation}
\label{2.3.8}
dp_\varphi=\int \sin\xi dp_\xi=
\frac{dmv}{2\pi}\cdot\frac{\pi}{2}\cdot \int_0^\pi\sin\xi d\xi=
\frac{dmv}{2},
\end{equation}  
or
\begin{equation}
\label{2.3.9}
d\mathbf p=\frac{dmv}{2}\mathbf n_\varphi,
\end{equation}  
where $\mathbf n_\varphi$ is the unit vector along the $\varphi$ 
axis.

Thus, the momentum of the element $dm$ propagating as a wave into 
a solid angle of $2\pi$ is oriented along the $\varphi$ axis. 
The momentum has no other components. This momentum is equal 
in magnitude to one half of the momentum the same element 
would have if it moved as a whole in one direction.

We can now calculate trivially the angular momentum of the element 
$dm$. Using the conventional expression for determination of the 
angular momentum $\mathbf M=[\mathbf r\times\mathbf p]$, and 
bearing in mind that the resultant momentum of the element $dm$ 
propagating as a wave is directed along the $\varphi$ axis, i.e., 
perpendicular to the $C$ plane, we arrive at the following 
expression for the resultant angular momentum of the element $dm$:
\begin{equation}
\label{2.3.10}
d\mathbf M_\Omega=\frac{dmvr}{2}[\mathbf n_r\times\mathbf n_\varphi],
\end{equation}  
where  $\mathbf n_r$  is the unit vector along the $r$ axis. We 
are not using here the relation for the angular momentum in its 
conventional form $d\mathbf M=dm[\mathbf r\times\mathbf v]$, 
because the velocity $v$ of an element propagating into a 
hemisphere has not specific direction. As evident from Eq. 
(\ref{2.3.10}), the angular momentum $d\mathbf M_\Omega$ lies 
in the plane $C$ (see Fig. 2.10).

\begin{figure}
\label{F10}
\begin{center}
\includegraphics[scale=0.2]{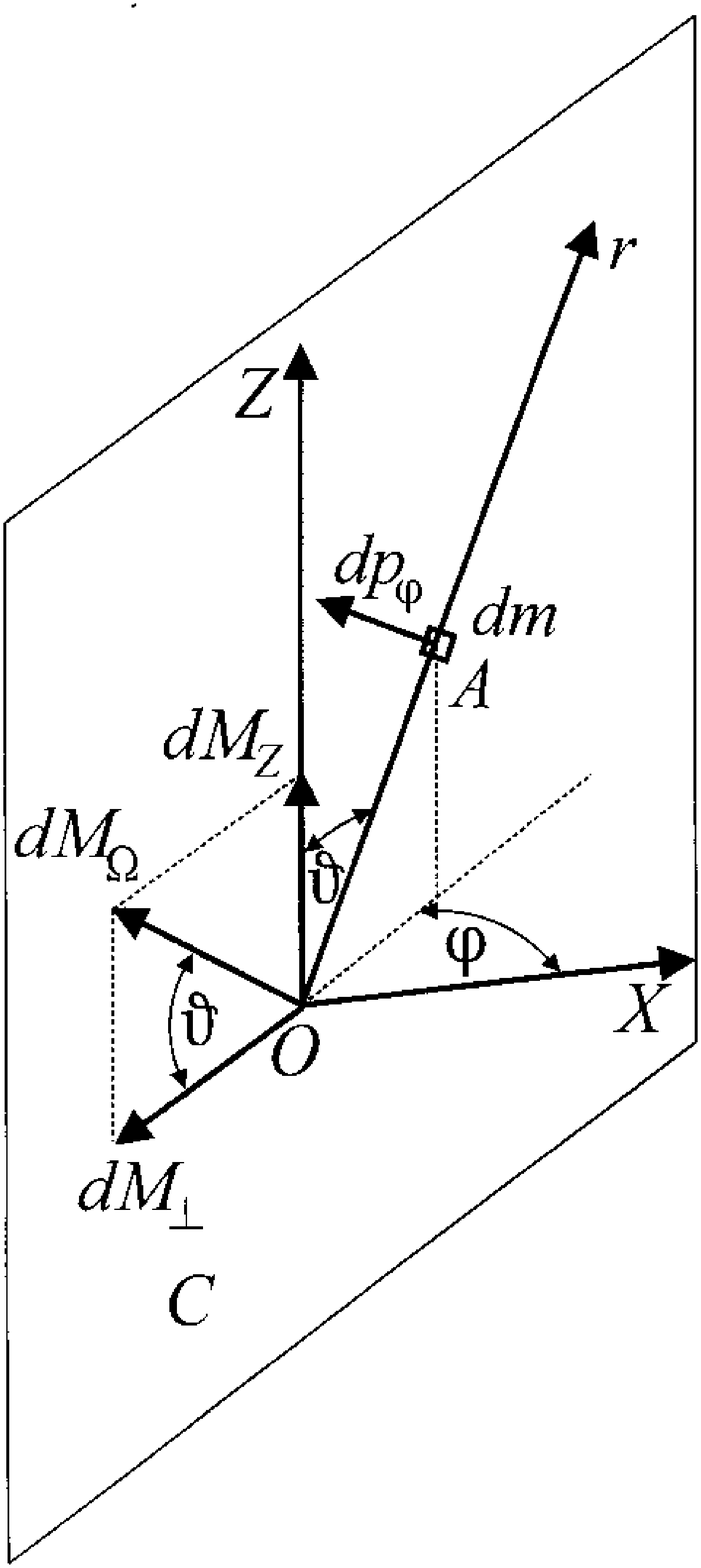}
\caption{}
\end{center}
\end{figure}

Compare now Eq. (\ref{2.3.10}) with the standard expression 
for the angular momentum of an element moving as a whole along 
a circular trajectory of radius $r$ and see what orbit could 
be identified with motion of the element as a wave. To avoid 
confusion, we shall denote the element of mass $dm$ by 
$dm_b$ $(dm_{body})$ in the case where we shall believe the 
element to move as a whole. It should be stressed, however, 
that $dm$ and $dm_b$ are one and the same element. Because 
the angular momentum of an element in a Coulomb potential is 
conserved, one can calculate the momentum at any point we 
choose. Let us calculate the angular momentum of the element 
$dm_b$ at the point of the circular orbit in which this element 
is farthest from the equatorial plane. At this point, the 
velocity of the element is directed along the $\varphi$ axis. 
The angular momentum $d\mathbf M_b$ of this element can be 
written as
\begin{equation}
\label{2.3.11}
d\mathbf M_b=dm_b[\mathbf r\times\mathbf v]=
dm_bvr[\mathbf n_r\times\mathbf n_\varphi].
\end{equation}  
A cursory inspection of Eqs. (\ref{2.3.10}) and (\ref{2.3.11}) 
reveals that the angular momenta have the same orientation, 
while in magnitude the angular momentum of the element 
propagating into the hemisphere is one half only of that of 
the element moving along a circular orbit. Significantly, $r$ 
for the element propagating into the hemisphere is the distance 
from the element to the nucleus, while for the element in 
circular motion, $r$ is not only the distance from the element 
to the nucleus but the radius of the circular orbit as well, 
with the orbit located such that the element under consideration 
is at the point of the orbit farthest from the equatorial plane. 
Thus, to calculate the angular momentum of the element 
propagating into a solid angle, one can restrict oneself to 
finding that of the element rotating in the corresponding 
orbit and taking one half of it. We shall use subsequently 
this observation.

As seen from Eqs. (\ref{2.3.10}) and (\ref{2.3.11}), both 
the $d\mathbf M_\Omega$ and $d\mathbf M_b$ angular momenta 
lie in the $C$ plane (note that each element has its own $C$ plane).

Both in Eq. (\ref{2.3.10}) and (\ref{2.3.11}), velocity $v$ is 
the velocity of motion of an individual element $dm$. Significantly, 
the velocity of motion of an element does not coincide with that 
of the wave process (in propagation of the same element into a 
solid angle). Examining the expression for the momentum 
(\ref{2.3.9}), we see that the velocity of the wave process 
at a given point is one half the velocity of motion of elements 
at the same point, and it is directed along the $\varphi$ axis, 
whereas the velocities of the elements have the same magnitude 
but are differently directed within the solid angle of $2\pi$. 
That an element moves not as a whole but propagates rather into 
a solid angle is accounted for in the expression for the angular 
momentum through the coefficient 1/2 (compare the Eqs. 
(\ref{2.3.10}) and (\ref{2.3.11}) for circular motion of the element).

Expand the angular momentum of the element $dm$ into two components, 
along and perpendicular to the $Z$ axis (Fig. 2.10). The axial $Z$ 
component can be written as
\begin{equation}
\label{2.3.12}
dM_Z=dM_\Omega\cdot\sin\vartheta=\frac{dmvr}{2}\sin\vartheta, 
\end{equation}
where $\vartheta$ is the coordinate of point $A$ (see Figs. 2.8 and 
2.10). For the component perpendicular to the $Z$ axis we obtain
\begin{equation}
\label{2.3.13}
dM_{\perp}=dM_\Omega\cdot\cos\vartheta=\frac{dmvr}{2}\cos\vartheta. 
\end{equation}

Let us calculate the angular momentum of the atom as a whole. 
This can be done by summing up the angular momenta of all the 
elements. By virtue of the axial symmetry of the system, the 
 $M_{\perp}$ component vanishes to leave the $M_Z$ one only:
 \begin{equation}
\label{2.3.14}
M_Z=\int\frac{1}{2}vr\sin\vartheta dm=\int\frac{1}{2}mvr\sin\vartheta dV. 
\end{equation}
In this expression, one has to substitute for velocity $v$ the 
velocity of the element $dm$ moving circularly over the circle, 
i.e., actually  $dm_b$. This was substantiated by us earlier in 
a comparison of Eqs. (\ref{2.3.10}) and (\ref{2.3.11}).

The velocity of a circularly rotating element can be taken from 
Eq. (\ref{1.4.18}), because this equation describes the motion of 
an element $dm$ over a circle of radius $r=a\tau$. In this formula, 
however, one has to drop index $\varphi$, because circular motion 
of an element with which we compare propagation of an element into 
the hemisphere can occur in any plane passing through the nucleus.

As already mentioned, the circular orbit which is opposed to 
propagation of an element into a solid angle should be positioned 
such that the element under consideration is at the point farthest 
from the equatorial plane (it is in this case that the angular 
momentum $d\mathbf M_b$ will be confined to the $C$ plane). This 
provides a mutual one-to-one correspondence between the position 
of an element and of the orbit under consideration (excluding the 
elements at the equator). Indeed, a given element can be crossed 
by a multitude of other orbits, having different angles of tilt. 
In all of these orbits, however, the point farthest from the 
equatorial plane will not coincide with the element we are 
considering. Therefore, these orbits will contribute to other 
elements. This leaves only one orbit for the given element. Hence,
 in such a consideration each element will be identified with one
 and the only orbit, and in integration over elements we will not
 be plagued by the danger of taking some orbits more than once 
 into account.

As for an element being capable of moving along any tilted orbit, 
this has already been taken into account in deriving the expression 
(\ref{2.3.10}) for the angular momentum $d\mathbf M_\Omega$, where 
the coefficient 1/2 was obtained.

Equation (\ref{2.3.14}) has a fairly simple structure. It can be 
revealed readily by considering an equivalent, circularly rotating 
element $dm_b$. The factor $\sin\vartheta$ accounts for the tilt 
of the trajectory of the equivalent circularly rotating element, 
and coefficient 1/2, for the real element propagating into a solid 
angle rather than rotating circularly. The other terms of Eq. 
(\ref{2.3.14}) make up the standard expression for the angular momentum.

Substituting the expressions for velocity (\ref{1.4.18}) and for 
the density of mass, (\ref{2.1.11}) or (\ref{2.1.12}), into Eq. 
(\ref{2.3.14}), we come to
\begin{equation}
\label{2.3.20}
M'_Z=\frac{1}{2}\int\limits_0^\infty \frac{4m_e}{a^3\pi^2}e^{-2\tau}
 \frac{\alpha c}{\sqrt{\tau}}a^4\tau^3d\tau \int\limits_0^\pi \sin^3 
 \vartheta d\vartheta \int\limits_0^{2\pi} d\varphi ,
\end{equation}

\begin{equation}
\label{2.3.21}
M''_Z=\frac{1}{2}\int\limits_0^\infty \frac{3m_e}{a^3 2\pi}e^{-2\tau}
 \frac{\alpha c}{\sqrt{\tau}}a^4\tau^3d\tau \int\limits_0^\pi \sin^4 
 \vartheta d\vartheta \int\limits_0^{2\pi} d\varphi .
\end{equation}

Our calculations finally yield
\begin{equation}
\label{2.3.22}
M'_Z=\hbar\cdot0.499,\qquad M''_Z=\hbar\cdot0.519.
\end{equation}

The first value fits better the available experimental data. At 
the same time, we will not yet reject the second value which 
corresponds to the charge distribution expressed in terms of 
spherical functions.

We have to bear in mind the following points. First, these 
values were obtained in nonrelativistic approximation. Second, 
by nonrelativistic quantum mechanics, the moment of the ground 
state is zero, which is in conflict with experiment.

We are turning now to the virial theorem (\ref{1.4.15}). 
Strictly speaking, we had no grounds for applying this equation 
in the first Chapter of the Paper. The virial theorem in the 
form of Eq. (\ref{1.4.15}) was used to describe the motion of 
an element of mass/charge in a Coulomb potential well. In the 
first part of the Paper, however, an additional statement was 
tacitly introduced that the velocity of an element $v_\varphi$ 
is directed along the $\varphi$ axis only. This is in clear 
conflict with the conditions under which the theorem can be 
applied. The above reasoning and the assumption that charges 
can interpenetrate lift this additional statement. Charges 
can move in a Coulomb field along any allowed orbit. In this 
case, the virial theorem in the form of Eq. (\ref{1.4.15}) 
is certainly applicable to charges in a Coulomb potential well.

On these grounds one can forward one more comment. In the 
analysis of Eq. (\ref{1.1.5}) it was pointed out, in particular, 
that it resembles in form the Schr\"odinger equation, the 
only difference being that it contains a coefficient 4, 
whereas in the Schr\"odinger equation the coefficient is 2. 
We use Eq. (\ref{1.1.5}) to derive the charge distribution 
and, hence, the {\itshape\bfseries potential} energy of an 
electron. This potential energy coincides with the 
eigenvalue of Eq. (\ref{1.1.5}). If we substituted 
coefficient 2 in place of 4 in Eq. (\ref{1.1.5}) and calculated 
the energy as eigenvalues of this new Eq. (\ref{1.1.5}), 
we would have obtained the numerical value of the 
{\itshape\bfseries total} electron energy. A decrease of 
the coefficient by a factor two brings about a corresponding 
decrease of the calculated energy to one half. In actual fact, 
this is simply the result of our having tacitly added to 
Eq. (\ref{1.1.5}) the virial theorem, because by this theorem 
the total energy of the electron is one half of its potential 
energy. While this is acceptable if our goal is to numerically 
calculate the values we are interested in, straightforward 
logic suggests that Eq. (\ref{1.1.5}) should have the 
coefficient 4, because it is the electron {\itshape\bfseries 
potential} energy that we calculate with this equation.

\section{Comment on the absence of emission from a stationary 
orbit}

This is a short comment, but we believe it to be important 
enough to be presented in a separate paragraph. By quantum 
mechanics, a point charge rotates around a nucleus. (More 
precisely, an electron is in a state having a definite 
energy and a definite projection of its angular momentum on 
the $Z$ axis). In the frame of electrodynamics, a rotating 
electron should emit radiation, lose energy and eventually 
fall on the nucleus. In actual fact this just does not happen. 
To provide a proper explanation for this, it was assumed that 
the electron does not radiate when in a stationary orbit. 
Actually, this is one of the {\itshape\bfseries postulates} 
of quantum mechanics. The stationary orbit (more precisely, 
the steady state) is calculated by solving the Schr\"odinger 
equation. Considered in the frame of mathematics, there is 
nothing that could be questioned; indeed, if, by Schr\"odinger 
equation, there is a solution within which an electron is in 
steady state with a certain energy, hence, this energy does 
not change, and, hence, the electron will not radiate. Viewed 
in the physical context, however, it is not clear in what does 
a stationary electron orbit differ from the non-stationary one. 
Why an electron residing in one orbit does radiate, and in 
another one, does not?

The above assumption of the existence of a {\itshape\bfseries 
distributed} electron charge permits one to {\itshape\bfseries 
lift this postulate}. Indeed, each element of mass/charge moves 
in accordance with the laws of theoretical mechanics and 
electrodynamics. Each individual element rotating about a 
nucleus is involved in periodic motion and, thus, has to 
radiate. But the elements make up a distributed charge. Now 
motion of a distributed charge is no longer periodic. Although 
each element moves along its own separate trajectory, the 
motion of the distributed charge as a whole is actually a 
common circular motion of the total charge. Radiation of 
variable fields by one element is canceled by that of the 
other elements.

Significantly, overall motion of the distributed charge as 
a whole (which now is no longer periodic) generates a 
magnetic field. This is reflected in the atom having a 
magnetic moment. An analog of such motion could be a set 
of closed currents which are known not to radiate periodic 
fields while having a constant magnetic field.

\section{Conclusion}

The time has come for summing up the outcome of our reasoning.

In the first part of the Paper, we have put forward an 
assumption that electron in an atom, rather than being a 
point object, is a {\itshape\bfseries distributed charge}. 
Taken as a whole, this charge should be equal to that of 
the electron; therefore, in all cases there should exist 
in the atom the $S$ {\itshape\bfseries-state} of the 
distributed charge. We forwarded the {\itshape\bfseries 
equation of wave mechanics}, i.e., the equation which a 
charge distribution should obey. The solutions of this 
equation derived for the Coulomb potential of the nucleus 
identified the shape of the charge distribution in an atom. 
These solutions were normalized against the {\itshape\bfseries 
electron} charge (not by unity). We further invoked standard 
methods in use in electrostatics to derive the values of the 
{\itshape\bfseries potential energy} of interaction of a 
distributed charge with the nucleus for three states of the 
hydrogen atom, which were found to coincide with those well 
known from quantum mechanics. More than that, these values 
coincide with the eigenvalues of the equation of wave mechanics. 
Standard methods employed in electrostatics were used to find 
the {\itshape\bfseries fields and potentials } of distributed 
charges and to calculate again the energies of interaction of 
distributed charges with a nucleus.

This has led to another assumption that the electron represents 
actually not only a distributed charge but a {\itshape\bfseries 
distributed mass} as well. The distributed mass assumes a 
distribution of the same shape as the distributed charge, and 
its motion coincides with that of a distributed charge. It thus 
turns out that the electron is a distributed {\itshape\bfseries 
charge/mass object}. The distributed mass should be normalized 
by the {\itshape\bfseries electron mass} (rather than by unity). 
An analysis of various versions of motion of distributed charge 
and mass led to the conclusion that the velocity of the elements 
of charge/mass should {\itshape\bfseries increase} as they 
{\itshape\bfseries approach} the nucleus (if the charge of an 
electron behaved as a solid body, the velocity of the motion 
of its elements should be {\itshape\bfseries increasing with 
distance} from the nucleus). An assumption was made concerning 
the velocity distribution for the charge/mass distributions 
found, thus making it possible to calculate the kinetic and 
total energies for three states of the hydrogen atom. These 
values were demonstrated to coincide with those known from 
quantum mechanics.

Next, the same velocity distributions were used to calculate 
the angular momenta for the same states of the hydrogen atom. 
It was found that the angular momenta {\itshape\bfseries do 
not coincide} with the values known from experiment. This 
suggested that the atom does not possibly possess spherical 
symmetry.

In the second part of the Paper, two versions of a 
{\itshape\bfseries non-spherical} charge distribution in an 
atom were advanced. It was shown that while the charge is 
certainly non-spherical, the potentials of these charges are 
close to spherical. Said otherwise, such an atom would look as 
spherical to an observer. Moreover, as one comes closer to the 
nucleus, this potential approaches ever more nearly the Coulomb 
potential. These distributions were used to calculate the 
potential, kinetic, and total energies for three states of the 
hydrogen atom. All these values were demonstrated to coincide 
with those well known from quantum mechanics. The angular 
momenta did not, however, equate to reality. This initiated a 
deeper analysis of the behavior of elements of charge/mass in 
a Coulomb potential well.

By the laws of theoretical mechanics, elements of mass in a 
Coulomb potential well can move along not only circular but 
elliptical trajectories as well. In order for such motion to 
become realistic, however, two more suggestions had to be 
made. First: {\itshape\bfseries charges can interpenetrate 
one another} (as electromagnetic waves penetrate one through 
another). Second: if there are orbits in which an element 
possesses the same energies, this element can move in any 
of these orbits, and, more than that, {\itshape\bfseries 
simultaneously along all these orbits}. In other words, 
this is no longer the motion of individual elements; it is 
rather the {\itshape\bfseries motion of a wave}. This motion 
resembles the {\itshape\bfseries Huygens--Fresnel principle}, 
the only difference being that the trajectories of the 
elements obey the laws of theoretical mechanics and are 
in effect closed curves.

These assumptions formed a basis on which the angular 
momentum of the hydrogen atom in ground state was calculated, 
and {\itshape\bfseries was found to be equal} to $\hbar/2$, 
in excellent agreement with the experimental data. The 
experimental observation that the angular momentum of $S$ 
states is $\hbar/2$ was introduced into theoretical quantum 
mechanics as a postulate, as an intrinsic angular momentum 
of the electron (spin). (It may be reminded that in 
non-relativistic quantum mechanics the angular momentum of 
the $S$ states is zero.) Using this mechanism, it was found 
possible to {\itshape\bfseries calculate} the angular 
momentum of the ground state of the hydrogen atom drawing 
solely from the laws of {\itshape\bfseries theoretical 
mechanics}. 

Each element in an atom undergoes periodic motion along a 
circle or ellipse. The motion of all elements as a whole is, 
however, no longer periodic, and represents rather circular 
motion of the charge as a whole. There being no common 
periodic motion of the charges, the electron (distributed 
charge) residing in steady state {\itshape\bfseries should 
not radiate}. The only thing that exists is a constant 
magnetic field generated by a common circular motion of 
charges. For quantum mechanics, the statement that the 
electron in a stationary orbit does not radiate is essentially 
a postulate. Thus, {\itshape\bfseries this postulate can 
now be lifted}.

Now how could one visualize an atom containing an electron 
in the form of a distributed charge? The most pictorial way 
would possibly be to compare it with a drop of a liquid. 
Elements of the liquid within the drop move in different 
directions but, when summed, produce rotation of the drop 
as a whole. This rotation could be detected only by labeling 
somehow an element of the liquid. If the element is not 
labeled, rotation of the drop cannot be detected. In other 
words, an observer would believe this drop to be at rest. 
One should bear in mind, however, that the density of the 
drop increases toward the center, as does increase also the 
velocity of motion of the elements of the liquid.

\section{Appendix}

{\bfseries{Equations for calculation of the potential}} (the 
notation used is that of Sec. 2.2). Point of observation --
$\mathrm r$, point of integration -- $\mathrm r'$. 

\noindent
For $\mathrm r>\mathrm r'$:
$$
\mathrm{U}_Q(\mathrm{r,}\vartheta)=\sum_{l=0}^\infty 
\sqrt{\frac{4\pi}{2l+1}}\cdot \frac{Q_l({\mathrm r})
Y_l(\vartheta)}{\mathrm r^{l+1}},
$$
where the multipole moment $Q_l({\mathrm r})$:
$$
Q_l({\mathrm r})=\sqrt\frac{4\pi}{2l+1}\int_0^{\mathrm r}\rho
(\mathrm r',\vartheta')\mathrm r'^lY_l^*(\vartheta')dV'.
$$
For $\mathrm r<\mathrm r'$:
$$
\mathrm{U}_G(\mathrm{r,}\vartheta)=\sum_{l=0}^\infty 
\sqrt{\frac{4\pi}{2l+1}}\cdot {\mathrm r^l}G_l({\mathrm r})
Y_l(\vartheta),
$$
where the multipole moment $G_l({\mathrm r})$:
$$
G_l({\mathrm r})=\sqrt\frac{4\pi}{2l+1}\int_{\mathrm r}^\infty\frac{\rho
(\mathrm r',\vartheta')}{\mathrm r'^{l+1}}Y_l^*(\vartheta')dV'.
$$
Functions $Q_l({\mathrm r})$ and $G_l({\mathrm r})$ depend on 
$\mathrm r$ in the upper and lower limits of integration as on 
a parameter.

\noindent
{\bfseries{Spherical functions}}:

$$
Y_l (\vartheta)=\sqrt\frac{2l+1}{4\pi}P_l(\cos\vartheta)
$$
{\bfseries{Legendre polynomials}}:

$$
P_0=1
$$
$$
P_2(\cos\vartheta)=\frac{1}{2}\cdot(3\cos^2\vartheta-1)
$$
$$
P_4(\cos\vartheta)=\frac{1}{8}\cdot(35\cos^4\vartheta-
30\cos^2\vartheta+3)
$$
$$
P_6(\cos\vartheta)=\frac{1}{48}\cdot(63\cdot11\cdot\cos^6
\vartheta-15\cdot63\cdot\cos^4\vartheta+15\cdot21\cdot
\cos^2\vartheta-15)
$$
$$
P_8(\cos\vartheta)=\frac{1}{48\cdot8}\cdot(99\cdot13\cdot15
\cdot\cos^8\vartheta-99\cdot28\cdot13\cdot\cos^6\vartheta+
99\cdot30\cdot7\cdot\cos^4\vartheta-63\cdot15\cdot4\cdot
\cos^2\vartheta+15\cdot7)
$$
{\bfseries{Shape of distributed charge in this notation}}:
$$
\rho'_{1NS}=-\frac{4}{\pi^2}e^{-2\mathrm r}\sin\vartheta,\qquad
\rho''_{1NS}=-\frac{3}{2\pi}e^{-2\mathrm r}\sin^2\vartheta.
$$
The potential at a point $\mathrm r$ is a sum of potentials 
calculated for $\mathrm r>\mathrm r'$ and $\mathrm r<\mathrm r'$:
$\mathrm{U(r,}\vartheta)=\mathrm{U}_Q(\mathrm{r,}\vartheta)+
\mathrm{U}_G(\mathrm{r,}\vartheta)$.